%% file: main.tex
\documentclass[11pt]{article}
\input{preamble.sty}

\theoremstyle{plain}

\declaretheorem[name={Farkas' Lemma},numbered=no]{farkas_lemma}
\newcommand{\FarkasLem}{\hyperref[lem:farkas]{Farkas' Lemma}\xspace}

\theoremstyle{definition}
\declaretheorem[name={False Clause Search Problem},numbered=no]{CNF_Search}
\newcommand{\CNFSearch}{\hyperref[def:CNF_Search]{false clause search problem}\xspace}
\declaretheorem[name={Stabbing Planes},numbered=no]{SPdef}
\declaretheorem[name={$k$-DNF Resolution},numbered=no]{kDNFres}

\declaretheorem[name={Split Cut},numbered=no]{splitCut}
\declaretheorem[name={Resolution over Cutting Planes},numbered=no]{Reskdef}

\declaretheorem[name={Randomized Communication},numbered=no]{randomizedCC}

\declaretheorem[name={Nondeterministic Communication Complexity},numbered=no]{np_cc}
\declaretheorem[name={Deterministic Communication},numbered=no]{detCC}
\declaretheorem[name={Real Communication},numbered=no]{realCC}

\title{Stabbing Planes}
\author{Paul Beame, Noah Fleming, Russell Impagliazzo, Antonina Kolokolova, Denis Pankratov Toniann Pitassi, Robert Robere}
\begin{document}

\newgeometry{margin=1.4in,top=1.6in,bottom=1in}

\begin{center}
{\LARGE Stabbing Planes}
\\[1cm] \large

\setlength\tabcolsep{0em}
\newcommand{\myPad}{\hspace{2.9em}}
\begin{tabular}{c@{\myPad}c@{\myPad}c}
  Paul Beame$^\dagger$ &
	Noah Fleming$^\ddagger$ &
	Russell Impagliazzo \\[-.5mm]
  \small\slshape  University of Washington &
	\small\slshape Memorial University &
	\small\slshape UCSD\\[3mm]
  Antonina Kolokolova$^\ddagger$ &
	Denis Pankratov$^\ddagger$ &
	Toniann Pitassi$^\mathsection$ \\[-.5mm]
  \small\slshape Memorial University &
	\small\slshape Concordia University &
  \small\slshape Columbia University \& IAS \\[3mm]
  & Robert Robere$^\ddagger$ & \\[-.5mm]
  & \small\slshape McGill University &
\end{tabular}

\vspace{9mm}

\large
{\today}

\vspace{9mm}

\bf Abstract

\end{center}

\normalsize
\noindent

We develop a new semi-algebraic proof system called Stabbing Planes which formalizes modern branch-and-cut algorithms for integer programming and is in the style of DPLL-based modern SAT solvers. As with DPLL there is only a single rule: the current polytope can be subdivided by branching on an inequality and its ``integer negation.'' That is, we can (nondeterministically choose) a hyperplane $ax \geq b$ with integer coefficients, which partitions the polytope into three pieces: the points in the polytope satisfying $ax \geq b$, the points satisfying $ax \leq b-1$, and the middle slab $b - 1 <
ax < b$. Since the middle slab contains no integer points it can be safely discarded, and the algorithm proceeds recursively on the other two branches. Each path terminates when the current polytope is empty, which is polynomial-time checkable. Among our results, we show that Stabbing Planes can efficiently simulate the Cutting Planes proof system, and is equivalent to a tree-like variant of the $\RCP$ system of Kraj\'i\v cek \cite{Krajicek98}. As well, we show that it possesses short proofs of the canonical family of systems of $\mathbb{F}_2$-linear equations known as the Tseitin formulas.  Finally, we prove linear lower bounds on the rank of Stabbing Planes refutations by adapting lower bounds in communication complexity and use these bounds in order to show that Stabbing Planes proofs cannot be balanced. In doing so, we show that real communication protocols cannot be balanced and establish the first lower bound on the real communication complexity of the set disjointness function.

\renewcommand*{\thefootnote}{\fnsymbol{footnote}}
\footnotetext[2]{Research supported by NSF Grants No. CCF-152424 and CCF-2006359.}
\footnotetext[3]{Research supported by NSERC.}
\footnotetext[4]{Research supported by NSERC, NSF Grant No. CCF-1900460 and the IAS school of mathematics.}
\footnotetext{A preliminary version of this work appeared at the $9$th \emph{Innovations in Theoretical Computer Science} (ITCS).} 
\thispagestyle{empty}
\setcounter{page}{0}
\newpage
\restoregeometry

\section{Introduction}
\include{Intro.tex}

\section{Preliminaries}
\label{sec:SPdef}
\include{StabbingPlanesDefn.tex}

\section{Refutations of the Tseitin Formulas} 
\label{sec:Tseitin}
\include{TseitinUpperBound.tex}

\section{The Relationship Between Stabbing Planes and Cutting Planes}
\label{sec:relationship_SP_CP}

It is an interesting question how Stabbing Planes compares to Cutting Planes, the main proof system based on ideas from integer programming. By contrasting the two systems we see three major differences:
\begin{itemize}
	\item \emph{Top-down vs. Bottom-up.} Stabbing Planes is a \emph{top-down} proof system formed by performing queries on the polytope and recursing; while Cutting Planes is a \emph{bottom-up} proof system, formed by deducing new inequalities from previously deduced ones.
	\item \emph{Polytopes vs. Halfspaces.} Individual ``lines'' in a Stabbing Planes proof are polytopes, while individual ``lines'' in a Cutting Planes proof are halfspaces.
	\item \emph{Tree-like vs. dag-like.} The graphs underlying Stabbing Planes proofs are trees, while the graphs underlying Cutting Planes proofs are general dags: intuitively, this means that Cutting Planes proofs can ``re-use'' their intermediate steps, while Stabbing Planes proofs cannot.
\end{itemize}
When taken together, these facts suggest that Stabbing Planes and Cutting Planes could be incomparable in power, as polytopes are more expressive than halfspaces, while dag-like proofs offer the power of line-reuse. Going against this intuition, we show next that Stabbing Planes and Cutting Planes are in fact very closely related. 

\subsection{Stabbing Planes Simulates Cutting Planes}
\label{subsec:CP_Simulation}
\include{CPSimulation.tex}

\subsection{Towards a Topology Preserving Simulation of Cutting Planes}
\label{subsec:SP_topology_simulation_CP}

An artifact of both the simulations of $\CP$ by variants of $\SP$ and of $\SP^*$ by $\CP$ is that they are far from being \emph{depth-preserving}; they convert shallow proofs into ones that are \emph{extremely} deep. In this section, we explore whether this explosion in depth is inherent to the simulation of $\CP$ by $\SP$.
While we are unable to conclusively resolve this question --- indeed, at this time the only technique for proving super-logarithmic depth lower bounds on $\SP$ works equally well for $\CP$ --- we provide a number of depth-preserving simulations of subsystems of $\CP$. 

To motivate our results, we will take a detour and discuss the relationship between $\SP$, $\CP$ and real communication protocols. Presently, almost all known lower bounds for $\CP$ are obtained by studying the communication complexity of the \emph{false clause search problem} (defined in \autoref{sec:SPdef}). For instance, it is known that:
\begin{itemize}
	\item A depth $d$ $\CP$ refutation yields a $d$-round real communication protocol for the associated false clause search problem.
	\item A size $s$ $\tCP$ refutation yields a real communication protocol $O(\log s)$-round real communication protocol for the associated false clause search problem.
	\item A size $s$ and space $\ell$ $\CP$ refutation yields a $O(\ell \log s)$-round real communication protocol for the associated false clause search problem.
	\item A size $s$ $\CP$ proof yields a dag-like real communication protocol for the associated false clause search problem.
\end{itemize}
All of these results have been used to derive strong lower bounds on Cutting Planes by proving the corresponding lower bound against the false clause search problem \cite{BonetEGJ00,deRezendeNV21, FlemingPPR17, HrubesP17, Krajicek98, Pudlak97}. Furthermore, this technique applies even to the stronger \emph{semantic} $\CP$ system, 
as all one needs to exploit is that the lines are linear inequalities, rather than expoiting some weakness of the deduction rules. However, this strength also illustrates a weakness of current techniques, as once the lines of a proof system become expressive enough, proof techniques which work equally well for semantic proof systems break down since every tautology has a short semantic proof. Therefore, it is of key importance to develop techniques which truly exploit the ``syntax'' of proof systems, and not just the expressive power of the lines.

Hence, it is somewhat remarkable that we are able to show that each of the simulation results above still hold if we replace real communication protocols with $\SP$ refutations, which are syntactic objects. That is, we show
\begin{itemize}
	\item[(i)] 	A depth $d$ $\CP$ refutation yields a depth $2d$ $\SP$ refutation. 
	\item[(ii)] A size $s$ $\tCP$ refutation yields a size $O(s)$ and depth $O(\log s)$ $\SP$ refutation.
	\item[(iii)] A size $s$ and space $\ell$ $\CP$ refutation yields a size $O(2^\ell s)$ and depth  $O(\ell \log s)$ $\SP$ refutation.
	\item[(iv)] A size $s$ $\CP$ refutation yields a size $O(s)$ $\SP$ refutation
\end{itemize}

\subsubsection{Simulating $\CP$ Depth}
\include{SPsimCPDepth.tex}

\subsubsection{Balancing $\tCP$ Proofs into $\SP$}
\include{balancingtCPtoSP.tex}

\subsubsection{Balancing Low-Space $\CP$ Proofs into $\SP$}
\include{SpaceCPtoSP.tex}

\subsection{Simulating non-CG Cuts}
\label{sec:non_CG_cuts}
\include{non_cg_cuts.tex}

\section{Relationship Between Stabbing Planes and Other Proof Systems}
\label{sec:SP_relations_to_other_proofs}

Having explored in depth the relationship between Cutting Planes and Stabbing Planes, we now describe how Stabbing Planes relates to other proof systems. A summary of these relationships can be seen in \autoref{fig:spRelations}.

Let us first note some of the separations that have already been established. 
\begin{itemize}
	\item Lower bounds for unsatisfiable systems of linear equations over finite fields, which are known for \emph{Nullstellensatz}~\cite{Grigoriev98}, the \emph{Polynomial Calculus}~\cite{BussGIP01},  \emph{Sum-of-Squares}~\cite{Grigoriev01,Schoenebeck08}, \emph{$\AC^0$-Frege}~\cite{ Ben-Sasson02, PitassiRST16, Hastad17, GalesiIRS19}, rule out the possibility of these systems simulating $\CP$.
	\item G\"{o}\"{o}s et al.~\cite{GoosKRS19} gave an exponential separation between \emph{Nullstellensatz} and Cutting Planes by observing that, for any unsatisfiable system of linear equations $F$, composing with the $m$-bit \emph{index gadget} can only increase the degree of refuting $F$ in Nullstellensatz by $O(\log m)$. On the other hand, Garg et al.~\cite{GargGKS18} showed that composing any function which requires resolution refutations of \emph{width} $w$ when composed with the index gadget requires Cutting Planes proofs of size $n^{\Omega(w)}$. Thus, any function which requires large resolution width but small Nullstellensatz degree provides such a separation. 
	\item Semantic $\CP$ is not polynomially verifiable, and therefore, assuming $\P \neq \NP$, no propositional proof system can simulate it. Indeed, Filmus, Hrube\v{s}, and Lauria observed that it has $O(1)$ size refutations of unsatisfiable instances of the $\NP$-complete subset sum problem.  
\end{itemize}



We establish the remaining simulations and separations in \autoref{fig:spRelations} next.

\subsection{Equivalence Between Stabbing Planes and treelike R(CP)}

\begin{figure}
	\centering 
		\hspace{3em}\begin{tikzpicture}
		
 		\node[rectangle,draw=black!60, very thick, rounded corners=0.5ex,fill=green!9,minimum size=18pt] at (2,1)(SP){$\SP =\tRCP$};
 		
 		\node[rectangle,draw=black!60, very thick, rounded corners=0.5ex,fill=green!9,minimum size=18pt] at (-2,1)(SP){$\CP$};
 		\node[rectangle,draw=black!60, very thick, rounded corners=0.5ex,fill=green!9,minimum size=18pt] at (0,-0.5)(SP){$\tCP$};
 		\node[rectangle,draw=black!60, very thick, rounded corners=0.5ex,fill=green!9,minimum size=18pt] at (0,2.5)(SP){$\RCP$};
 		\draw[->,color=black!60, very thick] (0.7,-0.15) -- (2,0.6);
		\draw[->,color=black!60, very thick] (-0.7,-0.15) -- (-1.9,0.6);
		\draw[->,color=black!60, very thick] (-1.9,1.4) -- (-0.7,2.15);
		\draw[->,color=black!60, very thick] (1.9,1.4) -- (0.7,2.15);
		\draw[->,color=black!60, very thick] (-1.55,1) -- (0.5,1);
 	\end{tikzpicture}
	\caption{Relationships between $\RCP$, $\CP$, and their tree-like variants. 
	An arrow from proof system $P_1$ to $P_2$ indicates that $P_2$ can polynomially simulate $P_1$.} \label{fig:spRelationsCP}
\end{figure}

\label{sec:SPequivtRCP}
\include{SPEqualstRCP.tex}

\subsection{Stabbing Planes Simulates Tree-like DNF Resolution}
\include{SPsimResk.tex}

\section{Lower Bounds on Stabbing Planes}
\label{sec:towards_unrestricted_SP_bounds}
Next, we tackle the problem of proving lower bounds on Stabbing Planes proofs. First, we show that near-maximal depth lower bounds on unrestricted Stabbing Planes proofs can be obtained by a straightforward reduction to communication complexity. Next, while we are unable to prove unrestricted size lower bounds, we explain why current techniques that would attempt to leverage the depth lower bounds fail.
 In doing so, we show that real communication protocols cannot be balanced by establishing the first superlogarithmic lower bound on the real communication complexity of the set disjointness function. 
 
 First, we recall some standard models of communication and previous lower bounds on depth via communication complexity. In proving lower bounds on proof complexity it has been fruitful to study the following associated search problem, introduced by Lov\'{a}sz et al.~\cite{LovaszNNW95}.
 
 \begin{CNF_Search}
\label{def:CNF_Search}
	Let $F=C_1 \land \ldots \land C_m$ be a CNF formula and let $(X,Y)$ be any partition of its variables. The associated false clause search problem $\Search_F^{(X,Y)} \subseteq \{0,1\}^{|X|} \times \{0,1\}^{|Y|} \times [m]$ is defined as $(x,y,i) \in \Search_F^{(X,Y)}$ if and only if $C_i(x,y) = 0$.
\end{CNF_Search}

 Our depth lower bounds are inspired by the approach of Impgliazzo et al.~\cite{ImpagliazzoPU94} who observed that small treelike Cutting Planes proofs implied short protocols in certain models of communication for solving the following false clause search problem.
 
 \begin{detCC} A deterministic communication protocol for a search problem $\mathcal{S} \subseteq \mathcal{X} \times \mathcal{Y} \times \mathcal{O}$ consists of two players, Alice and Bob. They receive private inputs $x \in \mathcal{X}$ and $y \in \mathcal{Y}$ respectively, and their aim is to agree on some $o \in \mathcal{O}$ for which $(x,y,o) \in \mathcal{S}$. To do so, they are allowed to communicate by sending messages to each other (in the form of a single bit) according to some predetermined \emph{protocol}. 
 This can be modelled combinatorially: every step in the communication protocol is associated with \emph{rectangle} of inputs $\mathcal{X}' \times \mathcal{Y}' \subseteq \mathcal{X} \times \mathcal{Y}$ consistent with the communication thus far; $\mathcal{X}'$ models what Bob knows about Alice's input, and $\mathcal{Y}'$ models what Alice knows about Bob's. If Alice communicates a bit, then this partitions $\mathcal{X}'$ into $\mathcal{X}_0'$ and $\mathcal{X}_1'$ corresponding to whether the bit Alice sent was $0$ or $1$. The communication ends when $\mathcal{X}' \times \mathcal{Y}'$ is \emph{monochromatic}, meaning that there is some $o \in \mathcal{O}$ such that $(x,y,o) \in \mathcal{S}$ for every $(x,y) \in \mathcal{X}' \times \mathcal{Y}'$.
 
 The \emph{deterministic communication complexity}  of computing $\mathcal{S}$ is the minimum number of bits communicated, or \emph{rounds} of communication, needed to solve $\mathcal{S}$ on any input $(x,y) \in \mathcal{X} \times \mathcal{Y}$. 
\end{detCC}

Observe that for $x, y \in \{0,1\}^n$, any integer linear inequality $ax +by \geq d$ can be evaluated in $w= \log \|a\|_1+\log \|b\|_1$ bits by Alice communicating $ax$ to Bob, and Bob responding with $by$. Therefore, for example, a Cutting Planes refutation of a CNF formula $F$ of depth $d$ in which the size of the coefficients are at most $2^w$ implies a $O(dw)$-round communication protocol for solving $\Search_F^{(X,Y)}$ for any partition $(X,Y)$ of the variables. 
By strengthening the model of  communication, we can simulate arbitrary linear inequalities. Next, we define two models of communication which allow us to do this; the first is the standard model of randomized communication complexity. 

\begin{randomizedCC}
A (bounded error) randomized communication protocol solving a search problem $\mathcal{S} \subseteq \mathcal{X} \times \mathcal{Y} \times \mathcal{O}$ is a distribution over deterministic communication protocols such that for every $(x,y) \in \mathcal{X} \times \mathcal{Y}$, with probability at least $2/3$, the protocol outputs $o$ for which $(x,y,o) \in \mathcal{S}$. The randomized communication complexity of $\mathcal{S}$ is the minimum number of rounds of any randomized protocol computing $\mathcal{S}$, where the number of rounds of a randomized protocol is the maximum number of rounds of any protocol with non-zero support in the distribution. 
\end{randomizedCC}

An alternative model, which more directly simulates arbitrarily linear inequalities, is the \emph{real communication} model introduced by Kraj\'{i}\v{c}ek~\cite{Krajicek98}. 

\begin{realCC} In a \emph{real communication protocol} for a search problem $\mathcal{S} \subseteq \mathcal{X} \times \mathcal{Y} \times \mathcal{O}$, the players Alice and Bob communicate via a ``referee''. In each round, Alice and Bob send real numbers $r_A, r_B$ to the referee who responds with a single bit $b$ which is $1$ if $r_A > r_B$, and $0$ otherwise. The \emph{real communication complexity} of computing $\mathcal{S}$ is the number minimum number of rounds needed to communication needed to solve $\mathcal{S}$ on any input $(x,y) \in \mathcal{X} \times \mathcal{Y}$.	
\end{realCC}

\subsection{Depth Lower Bounds}
\label{sec:SPDepthBounds}
\include{SPDepth.tex}

\subsection{Barriers to Size Lower Bounds}
\label{subsec:SPbarriers}
\include{SPBarriers.tex}

\section{Conclusion}
\label{sec:SPConclusion}
\include{ConclusionSP.tex}

 \bibliographystyle{plain}
\bibliography{biblio}
\end{document}

%% file: Intro.tex
Proof complexity provides an effective and principled way to analyze classes of practical algorithms for solving $\NP$-hard problems.
The general idea is to formalize a class of algorithms as a proof system --- a set of sound rules for making logical inferences --- by extracting out the types of reasoning used in the algorithms. This allows us to discard the practical implementation details while still maintaining the techniques that the algorithm can employ. Thus, by proving lower bounds on the size of proofs in these proof systems, we obtain lower bounds on the runtime of the associated class of algorithms. 

An illustrative example is the \emph{DPLL} algorithm \cite{DavisLL62, DavisP60}, which forms the basis of modern conflict-driven clause learning algorithms for solving SAT. For a CNF formula $F$, the DPLL algorithm is the following recursive search algorithm for a satisfying assignment: choose a variable $x_i$ (non-deterministically, or via some heuristic), and then recurse on the formulas $F \restriction (x_i=0)$ and $F \restriction (x_i =1)$. If at any point a satisfying assignment is found, the algorithm halts and outputs the assignment. Otherwise, if the current partial assignment falsifies some clause $C$ of $F$, the recursive branch is terminated. If every recursive branch terminated with a falsified clause, then $F$ is unsatisfiable and we can take the recursion tree as a proof of this fact; in fact, such a DPLL tree is equivalent to a \emph{treelike resolution} refutation of $F$.

Beyond DPLL, the approach of using proof complexity for algorithm analysis has been successfully employed to study many other classes of algorithms. This includes \emph{Conflict-driven clause-learning} algorithms for SAT \cite{BayardoS97,MarquesS99,MoskewiczMZZM01}, which can be formalized using \emph{resolution} proofs \cite{DavisP60}; lift-and-project methods for solving integer programming problems, which are formalized by the \emph{Lov\'{a}sz-Schrijver}~\cite{LovaszS91}, \emph{Sherali-Adams}~\cite{SheraliA90}, and \emph{Sums-of-Squares proofs} \cite{Grigoriev01, BarakBHKSZ12} proof systems; and the classical \emph{cutting planes} methor for integer programming \cite{gomory1963algorithm,Chvatal73a}, which is formalized by \emph{Cutting Planes proofs} \cite{Chvatal73a, CookCT87,Chvatal84}.

In this work, we continue the study of algorithms for solving integer programming problems through the lens of proof complexity. 
Many classic $\NP$-optimization problems are naturally phrased using integer programming and, due to this, algorithms for integer programming have had a profound effect throughout computer science and beyond. Recall that in an integer programming problem, we are given a polytope $P \subseteq \reals^n$ and a vector $c \in \reals^n$, and our goal is to find a point $x \in P \cap \integers^n$ maximizing $c \cdot x$.  A classic approach for solving integer programming problems is to refine the polytope $P$ by introducing Chv\'{a}tal-Gomory (CG) \emph{cutting planes} \cite{gomory1963algorithm}. A CG cutting plane for $P$ is any inequality $ax \leq \lfloor b \rfloor$, where $a$ is an integral vector, $b$ is a rational vector, and every point in $P$ satisfies $ax \leq b$. Observe that a CG cutting plane removes non-integral points from $P$ while preserving integer points. Thus, with each additional cutting plane, the polytope becomes a better approximation to the integer hull. 

Chv\'{a}tal observed that the cutting planes method can be naturally formalized as a proof system. The \emph{Cutting Planes} proof system proves the integer-infeasibility of a polytope $P$ by deducing the empty polytope by a sequence of CG-cutting planes. This has led to a number of strong lower bounds on the runtime of these algorithms \cite{FlemingGIPRTW21, HrubesP17,Pudlak97, BonetPR97}, as well as bounds on related measures such as the Chv\'{a}tal rank (see for example \cite{BureshOppenheimGHMP06,GoosP18, ChvatalCH89}). However, while Cutting Planes has become a highly influential proof system in proof complexity, the original cutting planes algorithms suffer from numerical instabilities and difficulties choosing good heuristics, and are therefore seldom used on their own in practice. 

Modern algorithms for integer programming combine cutting planes with a branch-and-bound procedure \cite{LandD10, Balas65}, resulting in a class of optimization algorithms known as \emph{branch-and-cut algorithms} \cite{PadbergMG91}. These algorithms search for an integer solution to a polytope $P$ by the following two procedures
\begin{itemize}
	\item \emph{Branch.} Split $P$ into smaller polytopes $P_1,\ldots, P_k$ such that every integer solution to $P$ lies in at least one of $P_1,\ldots, P_k$. 
	\item \emph{Cut.} Refine $P_1,\ldots, P_k$ by introducing additional cutting planes. 
\end{itemize}
Finally, the algorithm recurses on each of the resulting polytopes. While this branching rule is extremely general, in practice branching is typically done by single variables: selecting a variable $x_i$ and branching on all possible integer outcomes $P \cap \{x_i = t\}$ for each feasible integer value $t$. 
Other schemes that have been used in practice include branching on the hamming weight of a subset of variables \cite{FischettiL03} or branching using basis-reduction techniques~\cite{AardalL04,KrishnamoorthyP09,AardalBHLS00, Lenstra83, GrotschelLS84, LovaszS92}\footnote{For an in-depth discussions on branch-and-bound and branch-and-cut we refer the reader to \cite{WolseyN99, conforti2014integer, ApplegateBCC06}.}

While these branch-and-cut algorithms are much more efficient in practice than the classical cutting planes methods they are no longer naturally modelled by Cutting Planes proofs.
In this work we introduce\footnote{Recently (and subsequent to the conference publication of this work \cite{BeameFIKPPR18}), the authors became aware of the invited chapter of Pudl\'{a}k~\cite{Pudlak99} for the Logic Colloquium`97, in which he states that an unpublished work of Chv\'atal defines the Stabbing Planes system and the simulation of Cutting Planes. We take this independent discovery as evidence that Stabbing Planes is a highly natural proof system, which deserves further exploration.} the \emph{Stabbing Planes} ($\SP$) proof system in order to properly model branch-and-cut solvers. 
Intuitively, Stabbing Planes has the same branching structure as DPLL, but generalizes branching on single variables to branching on integer linear inequalities. 

We formalize Stabbing Planes in stages as a generalization of DPLL. Recall that in the setting of integer programming we want to prove the integer-infeasibility of a system of integer linear inequalities $Ax \geq b$ over real-valued variables. Further, suppose for simplicity of exposition that the system encodes a CNF formula $F$. 
We can rephrase DPLL \emph{geometrically} to the setting of integer linear programming. 
Consider some DPLL refutation of $F$. Each time the DPLL tree branches on the $\{0,1\}$-value of a variable $x_i$, instead  branch on whether $x_i \leq 0$ or $x_i \geq 1$; because the encoding $Ax\geq b$ of $F$ includes axioms $x_i \geq 0$ and $x_i \leq 1$ this is equivalent. After this replacement, each node $v$ in the DPLL tree is naturally associated with a polytope $P_v$ of points satisfying $Ax \geq b$ and each of the inequalities labelling the root-to-$v$ path. Since we began with a DPLL refutation, it is clear that for any leaf $\ell$, the polytope $P_\ell$ associated with the leaf is empty, as any $\mathbb{Z}^n$-valued point would have survived each of the inequalities queried on some path in the tree and thus would exist in one of the polytopes at the leaves. 

\begin{figure}	
		\centering
		\begin{tikzpicture}[scale=1]

		\draw[ thick, ->, black!60] (0,0)  -- (1.5,-1);
		\draw[ thick, ->, black!60] (0,0)  -- (-1.5,-1);
		\filldraw[color=black!60, fill=betterYellow!25, thick] (0,0) circle (5pt);
		\node[text width=2cm] at (2.1,-0.3) {\scriptsize $x+y \geq 2$};
		\node[text width=2cm] at (-1.1,-0.3) {\scriptsize $x+y \leq 1$};
		\draw[ thick, dashed, ->, black!60] (1.65,-1.17)  -- (2.05,-1.61);
		\draw[ thick, dashed, ->, black!60] (1.65,-1.17)  -- (1.25,-1.61);
		\filldraw[color=black!60, fill=white,  thick] (1.65,-1.17) circle (5pt);

		\draw[ thick, ->, black!60] (-1.65,-1.1)  -- (-0.65,-2.1);
		\draw[ thick, ->, black!60] (-1.65,-1.1)  -- (-2.65,-2.1);
		\node[text width=2cm] at (-2.6,-1.5) {\scriptsize$x-y \leq 0$};
		\node[text width=2cm] at (0.3,-1.5) {\scriptsize$x-y \geq 1$};
		\filldraw[color=black!60, fill=red!25,  thick] (-1.65,-1.15) circle (5pt);

		\draw[ thick, dashed, ->, black!60] (-2.8,-2.25)  -- (-2.4,-2.7);
		\draw[ thick, dashed, ->, black!60] (-2.8,-2.25)  -- (-3.2,-2.7);
		\draw[ thick, dashed, ->, black!60] (-0.5,-2.25)  -- (-0.9,-2.7);
		\draw[ thick, dashed, ->, black!60] (-0.5,-2.25)  -- (-0.1,-2.7);
		\filldraw[color=black!60, fill=white, thick] (-0.5,-2.25) circle (5pt);
		\filldraw[color=black!60, fill=white, thick] (-2.8,-2.25) circle (5pt);
		
		\draw[color=black!60, dashed, thick] (9,2) --(9,-4);
		\draw[color=black!60, dashed, thick] (5,2) --(5,-4);
		\draw[color=black!60, dashed, thick] (4,1) --(10,1);
		\draw[color=black!60, dashed, thick] (4,-3) --(10,-3);
		
		\draw[black!60,  thick, fill=green!9] (5, 1) rectangle (9, -3);
		\filldraw[color=black!60, thick,fill=red!25] (9,-3) -- (9, 1) -- (5, -3) -- cycle;
		\filldraw[color=black!60, thick,fill=betterYellow!25] (5,1) -- (9, 1) -- (9, -3) -- cycle;
		
		\draw[thick, black!60] (4.8, 1.2) -- (9.2, -3.2);
		\draw[thick, black!60] (8.8,1.2) -- (9.2, 0.8); 
		\draw[thick, black!60] (7,-1) -- (4.8, -3.2);
		\draw[thick, black!60] (9,-3) -- (8.8,-3.2);
		\draw[->, thick, black!60] (9,1) --(9.2,1.2) node[at end, above] {\scriptsize $x + y \geq 2$};
		\draw[->, thick, black!60] (6.4,-0.4) --(6.2,-0.6) node[at end, below] {\scriptsize $x + y \leq 1$\phantom{asf232}};
		\draw[->, thick, black!60] (6,-2) --(5.8,-1.8) node[at end, above] {\scriptsize $x - y \leq 0$\phantom{.}};
		\draw[->, thick, black!60] (8.9,-3.1) --(9.1,-3.3) node[at end, below] {\scriptsize $x - y \geq 1$};
		 \end{tikzpicture}	
		 \caption{A partial stabbing planes proof (left) and its result on the unit square (right). The yellow and red areas are removed from the polytope (green), and we recurse on both sides.}
		 \label{fig:spProofExample}
	\end{figure}

The Stabbing Planes system is then the natural generalization of the previous object: at each step in the refutation an \emph{arbitrary} integer linear form $ax$ in the input variables is chosen and we recurse assuming that it is at least some integer $b$ or at most its \emph{integer negation} $b-1$. 
Observe that because $a$ and $b$ are integral, any $x^* \in \mathbb{Z}^n$ will satisfy at least one of the inequalities $(ax \leq b-1, ~ax \geq b)$, and so if the polytope at each leaf (again, obtained by intersecting the original system with the inequalities on the path to this leaf) is empty then we have certified that the original system has no integral solutions (cf.~\autoref{fig:spProofExample}). The queries in a Stabbing Planes proof correspond to what is known as branching on \emph{general disjunctions} or \emph{split disjunctions} \cite{CookKS90} in integer programming, and capture the majority of branching that is done in practice \cite{conforti2014integer}. One of the major advantages of Stabbing Planes is its simplicity: refutations are decision trees that query integer linear inequalities. 

Recall that branch-and-cut solvers combine Stabbing Planes-style branching with additional cutting planes in order to refine the search space. 
We show that a single branching step in a Stabbing Planes proof can actually simulate CG cuts to the polytope, and therefore Stabbing Planes can simulate the execution of a branch-and-cut solver.
A novel corollary is that Stabbing Planes can polynomially-simulate dag-like Cutting Planes despite being a tree-like refutation system.
This simulation was extended by \cite{BasuCSJ20} to show that Stabbing Planes can simulate \emph{disjunctive cuts} which capture the vast majority of cutting planes used in practice, 
including \emph{lift and project cuts} \cite{BalasCC93}, \emph{split cuts} \cite{CookKS90}, \emph{Gomory mixed integer} cuts \cite{Gomory60} and \emph{MIR cuts}. 

Beyond providing a theoretical model for branch and cut algorithms, we believe that the simplicity of Stabbing Planes proofs, as well as its closeness to DPLL, makes Stabbing Planes a better starting point for the analysis and development of search algorithms based on integer programming, such as \emph{pseudoboolean} SAT solvers, than established proof systems. 
From the perspective of SAT solving, even though $\tRes$ is equivalent to DPLL, it is the search point of view of DPLL that has led to major advances in SAT algorithms. A natural hypothesis is that it is much easier to invent useful heuristics in the language of query-based algorithms, as opposed to algorithms based on the deductive rules of resolution. Stabbing Planes offers similar benefits with respect to reasoning about inequalities. Furthermore, Stabbing Planes is a direct generalization of DPLL, and therefore we hope that the fine-tuned heuristics that have been developed for modern DPLL-based solvers can be lifted to algorithms based on Stabbing Planes.

\subsection*{Stabbing Planes in Proof Complexity}
Despite its simplicity, Stabbing Planes proofs are remarkably powerful. As a motivating example, we give simple, short (quasi-polynomial size) $\SP$ proofs of systems of $\mathbb{F}_2$ linear equations known as the \emph{Tseitin formulas}. These formulas are one of the canonical hard examples for many algebraic and semi-algebraic proof systems, including Nullstellensatz \cite{Grigoriev98}, Polynomial Calculus \cite{BussGIP01}, and Sum-of-Squares \cite{Grigoriev01,Schoenebeck08}. 

\begin{thm}
\label{thm:TseitinUBInformal}
	There are quasipolynomial size Stabbing Planes proofs of any instance of the Tseitin tautologies. 
\end{thm}

\begin{figure}
	\centering 
		\hspace{3em}\begin{tikzpicture}
		
 		\node[rectangle,draw=black!60, very thick, rounded corners=0.5ex,fill=green!9,minimum size=18pt] at (-2.2,8)(SP){$\SP =\tRCP$};
 		
 		\node[rectangle,draw=black!60, very thick, rounded corners=0.5ex,fill=green!9,minimum size=18pt] at (2,8)(sCP){Semantic $\CP$};
 
 		\node[rectangle,draw=black!60, very thick, rounded corners=0.5ex,fill=green!9,minimum size=18pt] at (2.,6.5)(CP){$\CP = $ Facelike $\SP$};

 		\node[rectangle,draw=black!60, very thick, rounded corners=0.5ex,fill=green!9,minimum size=18pt] at (-1.7,6.5)(SP*){$\SP^*$};

 		\node[rectangle,draw=black!60, very thick, rounded corners=0.5ex,fill=green!9,minimum size=18pt] at (0,5)(CP*){$\CP^*$};
 		
 		\node[rectangle,draw=black!60, very thick, rounded corners=0.5ex,fill=green!9,minimum size=18pt] at (0,3.5)(CP*){$\tCP$};
 		
 		\node[rectangle,draw=black!60, very thick, rounded corners=0.5ex,fill=green!9,minimum size=18pt] at (0,10)(RCP){$\RCP$};

 		\node[rectangle,draw=black!60, very thick, rounded corners=0.5ex,fill=green!9,minimum size=18pt] at (-3.2,5)(tResk){$\tRes(k)$};
 		
 		\node[rectangle,draw=black!60, very thick, rounded corners=0.5ex,fill=green!9,minimum size=18pt] at (6,6.5)(SOS){Sum of Squares};
 		
 		\node[rectangle,draw=black!60, very thick, rounded corners=0.5ex,fill=green!9,minimum size=18pt] at (6,5)(SA){Nullstellensatz};
 		
 		\node[rectangle,draw=black!60, very thick, rounded corners=0.5ex,fill=green!9,minimum size=18pt] at (-5,6.5)(SP*){$\AC^0$-Frege};


  		\draw[->, color=black!60, very thick] (-1.7,6.9) -- (-1.7,7.6);
  		\draw[->, color=black!60, very thick] (2.3,6.9) -- (2.3,7.6);
  		\draw[<-, color=red!45, dashed, very thick] (1.3,6.9) -- (1.3,7.6);
  		\draw[<-, color=red!45, very thick] (0.2,6.5) -- (-1.15,6.5);
  		\draw[->, color=black!60, very thick] (0.5,5.4) -- (1.5,6.1);
  		\draw[->, color=black!60, very thick] (-0.5,5.4) -- (-1.5,6.1);
  		\draw[->, color=black!60, very thick] (0.3,6.9) -- (-0.6,7.6);
  		
  		\draw[->, color=black!60, very thick] (0.7,3.9) -- (2.3,6.1);
  		
  		\draw[<-, color=red!45, dashed, very thick] (-0.6,8) -- (0.7,8);
  		\draw[->, color=black!60, very thick] (-1.7,8.4) -- (-0.5,9.6);
  		\draw[<-,color=black!60, very thick, dashed] (2.5,8.4) -- (0.6,9.8);
  		\draw[->,color=red!45, very thick, dashed] (1.8,8.4) -- (0.2,9.57);
  		\draw[->,color=red!45, very thick, dashed] (3.8,6.7) -- (4.4,6.7);
  		\draw[<-,color=red!45, very thick, dashed] (3.8,6.3) -- (5.,5.4);
  		
  		\draw[<-,color=black!60, very thick] (-1.8,6.1) -- (-2.5,5.5);
  		\draw[<-,color=black!60, very thick] (6,6.1) -- (6,5.4);
      
     	 \draw[<-,color=black!60, very thick] (-4.5,6.) -- (-3.9,5.5);
     	 
  		\draw[->,color=red!45, very thick, dashed] (-2.3,6.3) -- (-3.9,6.3);
  		\draw[<-,color=black!60, very thick, dashed] (-2.3,6.7) -- (-3.9,6.7);
 	\end{tikzpicture}
	\caption{Known relationships between relevant proof systems. A solid black (red) arrow from proof system $P_1$ to $P_2$ indicates that $P_2$ can polynomially (quasi-polynomially) simulate $P_1$. A black (red) dashed arrow from $P_1$ to $P_2$ indicates that $P_2$ cannot polynomially (quasi-polynomially) simulate $P_1$.} \label{fig:spRelations}
\end{figure}


We also 
explore the relationships between $\SP$ and other proof systems. This is summarized in \autoref{fig:spRelations}. Most notably, we show that $\SP$ is polynomially-equivalent to $\tRCP$, the tree-like variant of Kraj\'{i}\v{c}ek's $\RCP$ proof system \cite{Krajicek98}. 
This system can be thought of as a mutual generalization of resolution and Cutting Planes, in which the lines are disjunctions of integer linear inequalities, and we are allowed to apply Cutting Planes rules on the inequalities and resolution-style cuts on the disjunctions.
  \begin{restatable}{thm}{SPeqtRCP}
  \label{thm:SPEqualstRCP}
  	The proof systems $\SP$ and $\tRCP$ are polynomially equivalent.	
  \end{restatable}
We note that even though $\SP$ is equivalent to a system already in the literature, the new perspective provided by it is indeed enlightening as none of the results in this section were known for $\tRCP$ (including the simulation of $\CP$!).

The remainder of this work tackles the problem of proving superpolynomial lower bounds on the complexity of $\SP$ proofs. 
Although we are unable to establish size lower bounds, we prove nearly optimal lower bounds on the \emph{depth} of $\SP$ refutations, as well as explain why several natural approaches for proving size bounds fail. 
The depth of an $\SP$ proof --- the longest root to leaf path --- is a natural parameter that captures the parallelizability of proofs, and is closely related to \emph{rank} measures of  polytopes, which have been heavily studied in integer programming theory \cite{ChvatalCH89}. 

\begin{restatable}{thm}{SPdepthLB}
\label{thm:SPdepthLB}
	There exists a family of unsatisfiable CNF formulas $\{F_n\}$ for which any $\SP$ refutation requires depth $\Omega(n/\log^2 n)$
\end{restatable}

The proof of this theorem proceeds by showing that shallow $\SP$ proofs give rise to short randomized and real communication protocols for the \emph{false clause search problem}, and then appealing to known lower bounds for this problem. 
In many strong proof systems such as Frege and Extended Frege it is known that depth lower bounds imply size lower bounds  via so-called \emph{balancing theorems} \cite{KrajicekProofBook}.
However, we can show that $\SP$ proofs cannot be balanced.
More precisely, $\SP$ proofs of size $s$ do not imply the existence of proofs of size $\poly(s)$ and depth $\polylog (s)$. 

While $\SP$ proofs cannot be balanced, the real communication protocols that result from $\SP$ proofs preserve the topology of the $\SP$ proof, and therefore size lower bounds on $\SP$ would follow by showing instead that the real communication protocols themselves could be balanced. There is a precedent for this: both deterministic and randomized communication protocols \emph{can} be balanced, and lower bounds on $\tCP$ proofs were obtained by exploiting this fact \cite{ImpagliazzoPU94}. However, we can also show that real communication protocols cannot be balanced. Enroute, we establish the first superlogarithmic lower bound on the real communication complexity of the set disjointness function, a function which has been central to many of the lower bounds which exploit communication complexity. 

\subsection{Related and Subsequent Work}

\paragraph{Lower Bounds on Variable Branching.} 
Lower bounds for a number of branch-and-cut algorithms using \emph{variable-branching} --- meaning that they branch on the integer value of single variables, rather than arbitrary inequalities (i.e., DPLL which branches on $x_i \in \mathbb{Z}$) --- have previously been established. The first example of this was the lower bound of Jerslow~\cite{Jerslow74} on the number of queries made by branch-and-bound algorithms with variable branching. 
Cook and Hartman proved exponential lower bounds on the number of operations required to solve certain travelling salesman instances by branch-and-cut algorithms which use variable branching and Chv\'{a}tal-Gomory cuts \cite{CookH90}. However, their lower bound is exponential only in the number of variables, and not in the number of inequalities. 
The first truly exponential (in the encoding size of the instance) lower bound for branch-and-cut algorithms which use variable branching was established by Dash~\cite{Dash05} by extending the lower bound of Pudl\'{a}k~\cite{Pudlak97} for Cutting Planes proofs. 

\paragraph{Lower Bounds for treeR(CP).} 
Lower bounds on certain restrictions of $\tRCP$ were established in earlier works. Kraj\'{i}\v{c}ek~\cite{Krajicek98} proved superpolynomial lower bounds on the size of $\RCP$ proofs when both the \emph{width} of the clauses and the magnitude of the coefficients of every line are sufficiently bounded. Concretely, letting $w$ and $c$ be upper bounds on the width and coefficient size respectively, the lower bound that he obtained was $2^{n^{\Omega(1)}}/c^{w \log^2 n}$. For $\tRCP$, Kojevnikov~\cite{Kojevnikov07} improved this lower bound to $\exp(\Omega(\sqrt{n/w \log n}))$, removing the dependence on the coefficient size.
In \autoref{sec:SPequivtRCP} we prove a size-preserving simulation of $\SP$ by $\tRCP$ which translates depth $d$ $\SP$ proofs into width $d$ $\tRCP$ proofs. Therefore, Kojevnikov's result implies a superpolynomial lower bound on the size of $\SP$ proofs of depth $o(n/\log n)$. In \autoref{sec:SPDepthBounds} we exhibit a formula for which any $\SP$ refutation requires depth $\Omega(n/\log^2 n)$.

\paragraph{Subsequent Work.} Following the conference version of this work \cite{BeameFIKPPR18}, Dadush and Tiwari \cite{DadushT20} showed that the Stabbing Planes proofs of the Tseitin formulas could be efficiently translated into $\CP$. This refuted a long-standing conjecture that $\CP$ requires exponential size refutations of these formulas \cite{CookCT87}. In the same paper, they established a polynomial equivalence between the number of nodes and the \emph{size} of $\SP$ proofs (i.e., the number of bits needed to express the proof). 

Dadush and Tiwari also considered $\SP$ in the context of \emph{mixed integer programming} (MIP) and proved exponential lower bounds on $\SP$ in this setting. 
In this setting, you are given a polytope $P=\{(x,y) \in \mathbb{R}^{n_1} \times \mathbb{R}^{n_2}: Ax + By \geq b\}$, and you are searching for an integer solution to the $x$-variables and a real solution to the $y$-variables. 
In other words, rather than proving that $P \cap \mathbb{Z}^n = \emptyset$, instead you would like to prove that $P \cap \mathbb{Z}^{n_1} \cap \mathbb{R}^{n_2} = \emptyset$. In this case, $\SP$ queries involving $y$-variables are disallowed, as this would not be sound. As shown by Dadush and Tiwari, this restriction turns out to be enough to obtain quite simple proofs of intractability. First, they prove that any $\SP$ refutation of a certain system of $2^n$ many inequalities (encoding the \emph{complete unsatisfiable formula}) requires size $2^n/n$. Next, they show that this system of inequalities admits a $\poly(n)$-size MIP \emph{extended formulation}.  As $\SP$ cannot branch on the extension variables, refuting this extended formulation is identical refuting the complete unsatisfiable formula in $\SP$. Dey,  Dubey, and Molinaro~\cite{DeyDM21} extended this technique to prove lower bounds for a number of MIP instances for packing, set cover, the Travelling Salesman problem, and the cross polytope, even when Gaussian noise is added to the coefficients.
However, this technique crucially relies on the fact that for MIP problems, $\SP$ queries cannot involve the real-variables, and therefore it does not appear to be possible to extend this technique to prove lower bounds on pure integer programming problems (i.e. those with only integer-variables) such as standard encodings of CNF formulas. 

Fleming et al.~\cite{FlemingGIPRTW21} showed that any Stabbing Planes proof with quasipolynomially bounded coefficients ($\SP^*$) can be translated into a Cutting Planes proof with at most a quasipolynomial increase in size. This allowed them to lift the exponential lower bounds on Cutting Planes proofs \cite{Pudlak97,FlemingPPR17,HrubesP17,GargGKS18} to $\SP^*$, and even to $\SP$ proofs with coefficients of size $\exp(n^\delta)$ for some constant $\delta <1$. As well, using this connection, they generalized the result of Dadush and Tiwari to show that there are quasipolynomial-size Cutting Planes refutations of \emph{any} unsatisfiable system of linear equations over the finite field $\mathbb{F}_p$ for any prime $p$. To prove this simulation of $\SP^*$ by Cutting Planes, they characterized Cutting Planes as a subsystem of Stabbing Planes, which they called \emph{facelike} Stabbing Planes. Briefly, a facelike Stabbing Planes proof restricts Stabbing Planes queries to have one side of the query be a face of the current polytope. Then, they show that $\SP$ proofs can be made facelike with a blowup that is proportional to the size of the coefficients and the diameter of the polytope. 

Basu et al.~\cite{BasuCSJ20} showed that $\SP$ can simulate \emph{disjunctive cuts}, a result which we cover in more detail in \autoref{sec:non_CG_cuts}. Furthermore, they give an IP instance can be solved in size $O(1)$ in $\SP$ but requires $\poly(n)$ deductions using split cuts.
As well, they explore the effect that \emph{sparsity} ---the number of non-zero coordinates in each query --- has on branch-and-cut. Sparse queries can be thought of as an intermediate between full $\SP$ branching and variable branching. They provide an instance where the any branch-and-bound tree must be of exponential size if the sparsity is $o(n)$.



Recently, Dantchev et al.~\cite{DantchevGGM21} introduced several novel techniques for proving lower bounds on $\SP$ proofs by exploiting their geometric structure. In particular, they make use of the fact that for a polytope $P$, every point $x^* \in P$ must be contained within some slab of the $\SP$ proof. This allowed them to establish linear lower bounds on the size of $\SP$ proofs of the \emph{pigeonhole principle} as well as the \emph{Tseitin formulas}; because $\SP$ proofs are binary trees, this leads to a depth $\Omega(\log n)$ lower bound for both formulas.

\subsection{Organization}

We begin by formally defining the Stabbing Planes proof system in \autoref{sec:SPdef}. The Stabbing Planes refutations of the Tseitin formulas are given in \autoref{sec:Tseitin}.
 \autoref{sec:relationship_SP_CP} explores how Stabbing Planes relates to Cutting Planes: we show that Stabbing Planes can polynomially simulate Cutting Planes, and we explore whether a simulation of Cutting Planes by Stabbing Planes can preserve other parameters of the proof in \autoref{subsec:SP_topology_simulation_CP}. We end this section (in \autoref{sec:non_CG_cuts}) by observing that Stabbing Planes can simulate most other types of cutting planes that are used in integer programming.
In \autoref{sec:SP_relations_to_other_proofs} we explore how Stabbing Planes compares to other popular proof systems in the literature, and we prove the equivalence with $\tRCP$. In the final section (\autoref{sec:towards_unrestricted_SP_bounds}) we explore whether we can prove lower bounds on Stabbing Planes with unbounded coefficients. We prove unrestricted depth lower bounds in \autoref{sec:SPDepthBounds} and rule out several natural approaches that utilize communication complexity in \autoref{subsec:SPbarriers}.

%% file: StabbingPlanesDefn.tex
We begin by formally defining the Stabbing Planes proof system. For this, it will convenient to use the following combination of Farkas' Lemma with Carath\'{e}odory's Theorem, which we state next. 

\begin{farkas_lemma}\label{lem:farkas} 
	Let $A \in \mathbb{Q}^{m \times n}$ and $b \in \mathbb{Q}^m$. Then exactly one of the following holds:
	\begin{itemize}
		\item[(i)] There exists $x \in \mathbb{R}^n$ such that $Ax \leq b$.
		\item[(ii)]	There exists $y \in \mathbb{Z}^m$ with $y \geq 0$ such that $y^\top A =0$ and $y^\top b < 0$. Moreover, $y$ has at most $n+2$ non-zero coordinates.
	\end{itemize}

\end{farkas_lemma}
The ``moreover'' part in (ii) follows from Carath\'{e}odory's Theorem. Proofs of both Carath\'{e}odory's Theorem and Farkas' Lemma can be found in \cite{conforti2014integer}.

With this, we are ready to define the Stabbing Planes proof system.   The Stabbing Planes proof system is a proof system for refuting the existence of integer-solutions to systems of linear inequalities; we will call such systems \emph{unsatisfiable}.  

\begin{SPdef}
Let $Ax \geq b$ be an unsatisfiable system of linear inequalities.
A \emph{Stabbing Planes ($\SP$) refutation} of $Ax \geq b$ is a directed binary tree, $T$, where each edge is labelled with a linear integral inequality satisfying the following \emph{consistency conditions}:
\begin{itemize}
	\item \emph{Internal Nodes.} For any internal node $u$ of $T$, if the right outgoing edge of $u$ is labelled with $cx \geq d$, then the left outgoing edge is labelled with its \emph{integer negation} $cx \leq  d-1 $.
	\item \emph{Leaves.} Each leaf node $v$ of $T$ is labelled with a conic combination of inequalities in $F$ with inequalities along the path leading to $v$ that yields $0 \geq 1$ (provided by \FarkasLem).
\end{itemize}
For an internal node $u$ of $T$, the pair of inequalities $(cx \le d-1,~ cx \ge d)$ is called the \emph{query} corresponding to the node. Every node of $T$ has a polytope $P$ associated with it, where $P$ is the polytope defined by the intersection of the inequalities in $F$ together with the inequalities labelling the path from the root to this node. We will say that the polytope $P$ \emph{corresponds} to this node. 
That is, if $P$ is the polytope corresponding to a node $v$ which queries $(cx \le d-1,~ cx \ge d)$, then the polytopes of the children of $v$ are given by $P \cap \{x \in \mathbb{R}^n : cx \le d-1\}$ and $P \cap \{x \in \mathbb{R}^n : cx \ge d\}$. 
For readability, we will use the abbreviation $P \cap \{cx \geq d\}$ for $P \cap \{x \in \mathbb{R}^n : cx \geq d\}$.
\end{SPdef}

The \emph{slab} corresponding to the query is $\{ x^* \in \mathbb{R}^n \mid d-1 < cx^* < d \}$, which is the set of points ruled out by this query. The \emph{width} of the slab is the minimum distance between $cx \leq d-1$ and $cx \geq d$, which is $1/\|c\|_2$. 
This gives an intuitive geometric interpretation of $\SP$ refutations: at each step we remove a slab from the polytope and recurse on the resulting polytopes on both sides of the slab. The aim is to recursively cover the polytope with slabs until every feasible point has been removed. An example of this can be seen in \autoref{fig:spProofExample}, where the yellow and red areas are the slabs of the two queries. 

The \emph{size} of an $\SP$ refutation is the bit-length needed to encode a description of the entire proof tree, which, for CNF formulas as well as sufficiently bounded systems of inequalities, is polynomially equivalent to the number of queries  in the refutation. In particular, Dadush and Tiwari~\cite{DadushT20} prove the following.

\begin{prop}[Corollary 1.2 in \cite{DadushT20}] \label{prop:coefficientBound}
	Let $Ax \geq b$ be any unsatisfiable system of linear equations whose coefficients require $\ell$ bits to express, and let $s$ be the number of nodes in an $\SP$ refutation of $Ax \geq b$. Then there exists an $\SP$ refutation of size $s \ell n^6$.
\end{prop}
As well, the \emph{depth} (or \emph{rank}) of the refutation is the depth of the binary tree. The depth of refuting an unsatisfiable system of linear inequalities $Ax \geq b$, denoted $\depth_\SP(Ax \geq b)$, is the minimum depth of any $\SP$ refutation of $Ax \geq b$.
Observe that any unsatisfiable system of inequalities $Ax \geq b$ whose corresponding polytope is contained within the unit cube $[0,1]^n$ (this includes the encodings of all CNF formulas) has a trivial size $2^n$ and depth $n$ $\SP$ refutation by branching on $(x_i \leq 0,~ x_i \geq 1)$ for every $i \in [n]$.

We will be particularly interested in Stabbing Planes refutations of unsatisfiable CNF formulas. Given a CNF formula $F$ we can translate it into an equisatisfiable system of linear inequalities in the natural way. First, introduce the inequalities $0 \leq x_i \leq 1$ for every variable $x_i$. Second, for each clause $\bigvee_{i \in I} x_i \vee \bigvee_{j \in J} \neg x_j$ introduce the inequality 
\[ \sum_{i \in I}x_i + \sum_{j\in J} (1-x_j) \geq 1. \]
It is easy to see that the this system of inequalities will have no integer solutions if and only if the original formula $F$ was unsatisfiable. With this translation we consider Stabbing Planes refutations of CNF formulas $F$ to be refutations of the translation of $F$ into a system of linear inequalities. 

With the previous translation in hand we show that $\SP$ is indeed a propositional proof system as defined by Cook and Reckhow~\cite{CookR79}. 

\begin{prop}
	Stabbing Planes is sound, complete, and polynomially verifiable.	
\end{prop}
\begin{proof}
	Completeness follows immediately from the fact that $\SP$ simulates DPLL, which is itself a complete proof system. Soundness follows because each slab in an $\SP$ proof, corresponding to a query $(cx \leq d-1,~ cx \geq d)$, removes only non-integral points. Indeed, by the integrality of $c$ and $d$, any $x \in \mathbb{Z}^n$ satisfies either $cx \leq d-1$ or $cx \geq d$. 
	Finally, to see that $\SP$ proofs are polynomially verifiable, observe that we only need to verify that every query is of the form $(cx \leq d-1,~cx \geq d)$ for integral $c$ and $d$, and that each conic combination labelling the leaves evaluates to $0 \geq 1$.
\end{proof}


We will also be interested in the \emph{Cutting Planes} proof system, the first proof system to formalize a class of integer programming algorithms. 

\paragraph{Cutting Planes.} A \emph{Cutting Planes proof} ($\CP$) of an inequality $cx \geq d$ from a system of integer linear inequalities $Ax \geq b$ is a sequence of inequalities $\{c_i x \geq d_i\}_{i \in [s]}$ such that $c_s = c$, $d_s = d$, and each inequality $c_ix \geq d_i$ either belongs to $Ax \geq b$ or is deduced from earlier inequalities in the sequence by one of the following inference rules
\begin{itemize}
	\item \emph{Linear Combination.} From inequalities $c_ix \geq d_i, c_jx \geq d_j$, deduce any non-negative linear combination with integer coefficients.	
	\item \emph{Division.} From an inequality $c_ix \geq d_i$, if $t \in \mathbb{Z}$ divides all entries in $c_i$ then deduce $(c_i/t)x \geq \lceil d_i/t \rceil$.
\end{itemize}
The \emph{size} of a Cutting Planes derivation is the number of inequalities $s$, and is known to be equivalent up to a polynomial blow-up to the complexity of expressing the proof \cite{CookCT87}. It is useful to visualize the derivation as a directed acyclic graph, where the nodes are the inequalities in the derivation, and for each inference, there are arcs from the at-most-two inequalities from which it was derived. With this in mind, the \emph{depth} of a Cutting Planes derivation is the length of the longest root-to-leaf path in the dag.  
Finally, a Cutting Planes \emph{refutation} of a system of integer linear inequalities $Ax \geq b$ is a derivation of the trivially false inequality $-1 \geq 0$.

%% file: TseitinUpperBound.tex
As a motivating example, we show that Stabbing Planes has quasipolynomial size proofs of the \emph{Tseitin formulas}, proving \autoref{thm:TseitinUBInformal}. For any graph $G=(V,E)$ and any labelling $\ell :  V \rightarrow \{0,1\}$ of the vertices, the \emph{Tseitin formula} of $(G,\ell)$ is the following system of $\mathbb{F}_2$-linear equations: for each edge $e$ we introduce a variable $x_e$, and for each vertex $v$ we have an equation 
\[  \bigoplus_{u: uv \in E} x_{uv} = \ell(v), \]
asserting that the sum of edge variables incident to $v$ must agree with its label $\ell(v)$. It is not difficult to see that a Tseitin formula is unsatisfiable iff $\sum_{v \in V} \ell(v)$ is odd. If we denote by $\deg(G)$ the maximum degree of any vertex in $G$ then the Tseitin formula of $(G,\ell)$ can be encoded as a CNF formula with $|V| \cdot 2^{\deg(G)-1}$ many clauses. The next theorem shows that there is a quasi-polynomial size Stabbing Planes refutation of any Tseitin formula. 

\begin{thm}
\label{thm:Tseitin}
	For any Tseitin instance $(G,\ell)$ there is a Stabbing Planes refutation of size $2^{O(\log^2 n) + \deg(G)}$ and depth $\deg(G) \cdot \log^2 n$. 
\end{thm}

First we will introduce some notation. For $U \subseteq V$ let $\ell(U) = \oplus_{v \in U} \ell(v)$. As well, if $U,W \subseteq V$ then let $E[U,W]$ denote the set of edges with one endpoint in $U$ and the other in $W$, and let $E[U]$ denote the set of edges with one endpoint in $U$.

\begin{figure}
	\center
	\begin{tikzpicture}[scale=0.8]
		\draw[color=black!60, very thick,fill=green!9]  plot [smooth cycle, tension=0.7] coordinates {(-0.9,-1.3) (-1.3,1.5) (0.8,3.5) (5,3.2) (9,3) (10.5,0.1) (8.5,-2.7) (5,-3) (1.2,-3)};
		
		\draw[very thick, red!50]  (0.6,1.9) -- (0.9,1.45);
		\draw[very thick, red!50]  (0,-.2) -- (0.5,0);
		\draw[very thick, red!50]  (1,-1.35) -- (1.25,-0.9);
		\draw[very thick, red!50]  (3.33,-2.05) -- (3.4,-1.5);
		\draw[very thick, red!50]  (6.35,-1.85) -- (6.2,-1.35);
		\draw[very thick, red!50]  (8.15,-1.18) -- (7.8,-0.75);
		\draw[very thick, red!50]  (9.25,0.4) -- (8.7,0.5);
		\draw[very thick, red!50]  (8,2.3) -- (7.9,1.75);
		\draw[very thick, red!50]  (5.8,2.15) -- (5.9,1.6);
		\draw[very thick, red!50]  (2.9,2.25) -- (2.85,1.7);
		
		\draw[color=red!50, very thick,fill=betterYellow!25]  plot [smooth cycle, tension=0.7] coordinates {(0.2,0) (0.5,1.4) (2,2) (5,1.8) (8,2) (9,0.5) (7.5,-1.2) (4,-1.8) (1.3,-1.2)};
		
		\draw[color=blue!45, very thick]  plot [smooth, tension=1.4] coordinates {(4.8,1.8) (5,0)  (4.7, -1.8)};
		
		\draw[very thick, blue!45]  (4.8,0.95) -- (5.2,1);
		\draw[very thick, blue!45]  (4.8,-.2) -- (5.2,-0.2);
		\draw[very thick, blue!45]  (4.7,-1.2) -- (5.1,-1.25);

		
		\node[text width=2cm] at (2.2,-0.3) {$U_1$};
		\node[text width=2cm] at (8.6,-0.3) {$U_2$};
		\node[text width=2cm] at (6.3,-2.2) {$U$};
		\node[text width=2cm] at (6.4,0.3) {$a$};
		\node[text width=2cm] at (1.1,1.2) {$b$};
		\node[text width=2cm] at (10.1,-2.8) {$G$};
	\end{tikzpicture}
	\caption{A single round of the algorithm. Note that $\kappa_{U_1} = a+b$ and $K_{U_2} = a+(K_U - b)$.}
	\label{fig:Tseitin_alg}	
\end{figure}

\begin{proof}

The Stabbing Planes proof will implement the following recursive search algorithm for a violated constraint of the Tseitin instance. In each round we maintain a subset $U \subseteq V$ and an integer $\kappa_U \in \mathbb{N}$ representing the total value of the edges $E[U]$ leaving $U$. Over the algorithm, we maintain the invariant that $\ell(U) + \kappa_U$ is odd, which implies that there is a contradiction to the Tseitin instance inside of $U$.

Initially, set $U:=V$ and $\kappa_V = 0$, and note that the invariant holds since $\ell(V)$ is odd by definition. Then perform the following algorithm (see also \autoref{fig:Tseitin_alg}):
\begin{enumerate}
	\item Choose a balanced partition $U= U_1 \cup U_2$ (so $||U_1|-|U_2|| \leq 1$)
	\item Query the value of $a = \sum \limits_{e \in E[U_1,U_2]}x_e$ and $b = \sum \limits_{e \in E[U_1] \setminus E[U_1, U_2]} x_e$.
	\item The value of the edges leaving $U_1$ is $\kappa_{U_1} := a+b$ and the value of the edges leaving $U_2$ is $\kappa_{U_2} := a + (\kappa_U - b)$; so we recurse on the subset that maintains the invariant. 
\end{enumerate}
First we note that exactly one of the two subsets must maintain the invariant. This follows from the next short calculation:
\begin{align*} 
	\ell(U) + \kappa_U &= \ell(U_1) + \ell(U_2) + \kappa_U + 2a &(\bmod 2) \\
	&=(\ell(U_1) + a+b) + (\ell(U_2) + a + (\kappa_U - b)) &(\bmod 2) \\
	&= (\ell(U_1) + \kappa_{U_1}) + (\ell(U_2) + \kappa_{U_2}) &(\bmod 2)
\end{align*}
thus, since $\ell(U) + \kappa_U$ is odd, it follows that exactly one of the $\ell(U_1) + \kappa_{U_1}$ or $\ell(U_2) + \kappa_{U_2}$ must also be odd. Second, we note that the recursion ends when $|U| =1$, at which point we obtain an immediate contradiction between $\kappa_U$ and the equation corresponding to the single node inside $U$.

To implement this algorithm in Stabbing Planes, it suffices to show how to perform the queries in step 2, and how to deduce a contradiction when $|U|=1$; we begin with step 2. The first query can be performed by a binary tree with $|E[U_1,U_2| \leq n$ leaves, one corresponding to each possible query outcome. Internally, the tree queries all possible integer values for the sum (e.g. $(a \leq 0,~a \geq 1), (a \leq 1, ~a \geq 2), \ldots$). The second query can similarly be performed by a tree with $|E[U_1]|\leq n$ leaves. Since we choose a balanced partition in each step, the recursion terminates in at most $O(\log n)$ rounds --- thus, we have a tree with branching factor $O(n)$ and depth $O(\log n)$, yielding a size bound of $n^{O(\log n)}$. Furthermore, each query can be implemented in a tree of depth $O(\log n)$, and so the depth of the proof is $O(\log^2 n)$.

For the leaf-case, when $|U|=1$, let $u$ be the unique vertex in $U$. Stabbing Planes has deduced that 
\begin{align*}
	\sum_{v:uv \in E} x_{uv} &=\kappa_U \neq \ell(u) &(\bmod 2). 	
 \end{align*}
  This is a contradiction to the Tseitin axiom $\oplus_{v:uv \in E} x_{uv} = \ell(v)$. However, the Tseitin formula is presented to $\SP$ as a system of linear inequalities (encoding a CNF formula) which is equivalent to the Tseitin formulas (a system of $\mathbb{F}_2$ linear equations) over  integer solutions. Therefore, while there are no integer solutions satisfying  $\kappa_U = \sum_{v:uv \in E} x_{uv} =\ell(u)$, there could still be \emph{non-integer} solutions. To handle this, we simply force each of the $\deg(G)$ variables involved in this constraint to take integer values by sequentially querying the value of each variable one-by-one. That is, for each $\{x_{uv}: uv \in E\}$ we query $(x_{uv} \leq 0,~x_{uv} \geq 1)$, noting that we have axioms saying that $x_{e} \geq 0$ and $x_{e} \leq 1$ for every $e \in E$. 
   This can be done in a binary tree of height $\deg(G)$ with at most $2^{\deg(G)}$ leaves, where at each leaf we derive $0 \geq 1$.
\end{proof}

Together with the lower bounds of Buresh-Oppenheim et. al.~\cite{BureshOppenheimGHMP06} and Fleming et. al.~\cite{FlemingGIPRTW21} on the depth of Cutting Planes and semantic Cutting Planes proofs of the Tseitin formulas, \autoref{thm:TseitinUBInformal} provides an exponential separation in terms of \emph{depth} for these proof systems and Stabbing Planes.

%% file: CPSimulation.tex
A Cutting Planes proof is of a linear inequality $cx \geq d$ from polytope $P$, presented as a list of integer linear inequalities $\{a_ix \geq b_i\}$, is a sequence of inequalities $\{c_i x \geq d_i\}_{i \in [s]}$ such that the final inequality is $cx \geq d$, and each $c_i x \geq d_i$ is either one of the inequalities of $P$, or is deduced from previously derived inequalities by one of the following two deduction rules:
\begin{itemize}
	\item \emph{Conic Combination}. From inequalities $ax \geq b$, $cx \geq d$ deduce any nonnegative linear combination of these two inequalities with integer coefficients.
	\item \emph{Division}. From an inequality $ax \geq b$, if $d \in \mathbb{Z}$ with $d \geq 0$ divides all entries of $a$ then deduce $(a/d)x \geq \lceil b/d \rceil$. 
\end{itemize}
A Cutting Planes \emph{refutation} is a proof of the trivially false inequality $0 \geq 1$. 

Equivalently, we can view a Cutting Planes refutation as a sequence of polytopes $P=P_1,\ldots, P_s =\emptyset$ such that $P_i$ is obtained from $P_{i-1}$  by including an inequality which can be deduced from the inequalities of $P_{i-1}$ by one of the two rules of Cutting Planes. In integer programming, we obtain $P_i$ from $P_{i-1}$ by a \emph{Chv\'{a}tal-Gomory cut}.

\begin{thm}
	Stabbing Planes polynomially simulates Cutting Planes.	
\end{thm}
\begin{proof} 
	We will say that a polytope $P'$ can be deduced from $P$ by Stabbing Planes if there is a query $(ax \leq b-1,~ ax \geq b)$ such that $P \cap \{x: ax \leq b-1\} = \emptyset$ and $P \cap \{x: ax \geq b\} \subseteq P'$. 
	
	Let $P$ be an unsatisfiable polytope and $P=P_1,\ldots, P_s = \emptyset$ be a Cutting Planes refutation. To prove the theorem, we show that for each $i \in [s-1]$, $P_{i+1} = P_i \cap \{x: ax \geq b\}$ can be deduced from from $P_i$ in Stabbing Planes by the query $(ax \leq b-1,~ax \geq b)$. It remains to show that $P_i \cap \{x: ax \leq b-1\}=\emptyset$. There are two cases, depending on the rule used to derive $ax \geq b$ from $P_i$. 
	\begin{itemize}
		\item If $ax \geq b$ was derived by a conic combination of inequalities belonging to $P_i$, then $P_i \cap \{ax \leq b-1\}=\emptyset$ can be witnessed by adding the conic combination equalling $ax \geq b$ together with $ax \leq b-1$ to deduce $0 \geq 1$. 
		\item Otherwise,  $ax \geq b$ is derived by division, i.e., it is  $(c/t)x \geq \lceil c/t \rceil$ for some integer $t \geq 0$. Then $\lceil d/t \rceil \geq d/t$ and so $0 \geq 1$ is a conic combination of $(c/t)x \leq \lceil d/t \rceil-1$ and $cx \geq d$, witnessing that $P \cap \{x: (c/t)x \leq \lceil d/t \rceil-1 \} = \emptyset$. 
	\end{itemize}
\vspace{-2em}	
	\begin{figure}	
		\centering
		 \begin{tikzpicture}[scale=0.8]
			\node[circle,draw=black!60, fill=betterYellow!25, minimum size=15pt, thick] at (0,0)(v1){\scriptsize$P_1$};
			\node[circle,draw=black!60, fill=betterYellow!25, minimum size=15pt, thick] at (1,-1.5)(v2){\scriptsize$P_2$};
			\node[circle,draw=black!60, fill=betterYellow!25, minimum size=15pt, thick, minimum size=0.5cm, inner sep=0.1em] at (2.88,-4)(v3){\scriptsize $P_{s-1}$};
			\node[circle,draw=black!60, fill=red!15, minimum size=15pt, thick] at (4,-5.5)(v4){\scriptsize$\emptyset$};
			\node[circle,draw=black!60, fill=red!15, minimum size=15pt, thick] at (2,-5.5)(v5){\scriptsize$\emptyset$};
			\node[circle,draw=black!60, fill=red!15, minimum size=15pt, thick] at (-1,-1.5)(v6){\scriptsize$\emptyset$};
			\node[circle,draw=black!60, fill=red!15, minimum size=15pt, thick] at (0,-3)(v7){\scriptsize$\emptyset$};
			\draw[black!60, thick, ->] (v1)--(v2);
			\draw[black!60, thick, ->] (v1)--(v6);
			\draw[black!60, thick, ->] (v2)--(v7);
			\draw[black!60, thick, ->] (v2)--(1.9,-2.7);
			\draw[black!60, thick, dashed] (v2) -- (v3);
			\draw[black!60, thick, ->] (v3)--(v4);
			\draw[black!60, thick, ->] (v3)--(v5);
			
			\node[text width=2cm] at (2,-0.5) {\scriptsize$a_1 x \geq b_1$};
			\node[text width=2cm] at (-1.5,-0.5) {\scriptsize$a_1 x \leq b_1-1$};
			\node[text width=2cm] at (-0.5,-2.2) {\scriptsize$a_2 x \leq b_2-1$};
			\node[text width=2cm] at (3,-2.2) {\scriptsize$a_2 x \geq b_2$};
			\node[text width=2cm] at (4.9,-4.7) {\scriptsize$a_s x \geq b_s$};
			\node[text width=2cm] at (1.4,-4.7) {\scriptsize$a_s x \leq b_s-1$};
		 \end{tikzpicture}	
		 \caption{The Stabbing Planes refutation which results from translating a Cutting Planes refutation $P=P_1,\ldots, P_s = \emptyset$.}
		 \label{fig:SPsimCP}
	\end{figure}
\end{proof}

To see that Stabbing Planes can simulate Cutting Planes, we view each inequality

%% file: SPsimCPDepth.tex
First, we exhibit a depth-preserving simulation of $\CP$ by $\SP$, which establishes (i). Furthermore, if the proof is \emph{tree-like} then this simulation simultaneously preserves the size. 

\begin{thm}
\label{thm:SPsimCPdepth}
	$\depth_\CP(F) \geq 2 \cdot \depth_\SP(F)$. Moreover, for any $\tCP$ refutation of depth $d$ and size $s$ there is an $\SP$ refutation of depth $2d$ and size $O(s)$.
\end{thm}
\begin{proof}
	It is sufficient to prove the ``moreover'' part of the statement, since, by recursive doubling, any $\CP$ refutation can be converted into a $\tCP$ refutation where the depth remains the same. 
	
	Fix a $\tCP$ refutation of size $s$ and depth $d$, and let $G$ be the its underlying tree. We will construct an $\SP$ refutation of the same system of linear inequalities by proceeding from the root of $G$ to the leaves. In the process, we will keep track of a subtree $T$ of $G$, which we have left to simulate, and an associated ``current'' node $v$ of the $\SP$ refutation that we are constructing. Along the way, the following invariant will be maintained: at every recursive step $(T,v)$ with $T \neq G$, if the root of $T$ is labelled with the inequality $ax \geq b$, then the edge leading to $v$ in the $\SP$ refutation is labelled with $ax \leq b-1$. 
	
	Initially, $T=G$ and the $\SP$ refutation contains only a single root node $v$. Consider a recursive step $(T,v)$. We break into cases based on which rule was used to derive the root of $T$. 
	\begin{figure}	
		\centering
		\hspace{2em}
		 \begin{tikzpicture}[scale=1]

		\draw[color=black!60, very thick, rounded corners=0.5ex,fill=green!9] (-2.2,-.3) -- (-2.2,0.3) -- (0.9,0.3) -- (0.9,-0.3) -- cycle ;
		\node[text width=4cm] at (0,0) {\scriptsize$\lambda(a+c)x \geq \lambda(c+d)$};
		
		\draw[color=black!60, very thick, rounded corners=0.5ex,fill=green!9] (-3.2,1.7) -- (-3.2,2.3) -- (-1.9,2.3) -- (-1.9,1.7) -- cycle ;
		\node[text width=2cm] at (-2,2) {\scriptsize$ax \geq b$};
		\draw[thick, ->,color=black!60] (-2.4,1.65)  -- (-1.6,0.4);
		
		\draw[thick, ->,color=black!60,dashed] (-3.7,3.5)  -- (-2.9,2.4);
		\draw[thick, ->,color=black!60,dashed] (-1.4,3.5)  -- (-2.2,2.4);
		
		\draw[color=black!60, very thick, rounded corners=0.5ex,fill=green!9] (1.8,1.7) -- (1.8,2.3) -- (0.5,2.3) -- (0.5,1.7) -- cycle ;
		\node[text width=2cm] at (1.7,2) {\scriptsize$cx \geq d$};
		\draw[thick, ->,color=black!60] (1,1.65)  -- (0.3,0.4);
		
		\draw[thick, ->,color=black!60,dashed] (0,3.5)  -- (0.8,2.4);
		\draw[thick, ->,color=black!60,dashed] (2.3,3.5)  -- (1.5,2.4);
		
		 \end{tikzpicture}	
		 \begin{tikzpicture}[scale=1]

		\draw[thick, ->, dashed,color=black!60] (-0.5,0)  -- (0.5,-1);
		\draw[thick, ->,color=black!60] (-0.5,0)  -- (-1.5,-1);
		\filldraw[color=black!60, fill=betterYellow!25,thick] (-0.5,0) circle (5pt);
		\node[text width=3cm] at (-2.6,-0.3) {\scriptsize $\lambda(a+c)x  \leq \lambda(b+d)-1$};		
		
		\draw[thick, ->,color=black!60] (-1.65,-1.1)  -- (-0.65,-2.1);
		\draw[thick, ->,color=black!60] (-1.65,-1.1)  -- (-2.65,-2.1);
		\node[text width=2cm] at (-2.6,-1.5) {\scriptsize$ax \leq b-1$};
		\node[text width=2cm] at (0,-1.5) {\scriptsize$ax \geq b$};
		\filldraw[color=black!60, fill=betterYellow!25,thick] (-1.65,-1.15) circle (5pt);

		\draw[thick, ->,color=black!60] (-0.5,-2.25)  -- (-1.5,-3.25);
		\draw[thick, ->,color=black!60] (-0.5,-2.25)  -- (0.5,-3.25);
		\filldraw[color=black!60, fill=betterYellow!25,thick,thick] (-0.5,-2.25) circle (5pt);
		\node[text width=2cm] at (1.2,-2.7) {\scriptsize$cx \geq d$};
		\node[text width=2cm] at (-1.5,-2.7) {\scriptsize$cx \leq d-1$};
		
		\draw[thick, dashed, ->,color=black!60] (0.65,-3.4)  -- (1.05,-3.84);
		\draw[thick, dashed, ->,color=black!60] (0.65,-3.4)  -- (0.25,-3.84);
		\filldraw[color=black!60, fill=betterYellow!25,thick,thick] (0.65,-3.4) circle (5pt);
		\node[text width=2cm] at (1.9,-3.4) {\scriptsize$\ell_3$};
		
		\node[circle,draw=black!60, fill=red!25, thick, inner sep =0.1em, minimum size=0.8em] at (-2.84,-2.3)(v9){\small$\emptyset$};
		
		\node[text width=2cm] at (-2.4,-2.26) {\scriptsize$\ell_1$};
		
		\node[circle,draw=black!60, fill=red!25, thick, inner sep =0.1em, minimum size=0.8em] at (-1.65,-3.45)(v9){\small$\emptyset$};
		\node[text width=2cm] at (-1.2,-3.4) {\scriptsize$\ell_2$};

		 \end{tikzpicture}	
		 \caption{A $\tRCP$ refutation invoking the conic combination rule (left) and the corresponding partial $\SP$ refutation (right).}
		 \label{fig:SPtCP}
	\end{figure}

	\begin{itemize}
		\item \emph{Conic Combination.} Suppose that the root of $T$ is labelled with an inequality $\lambda(a+c)x \geq \lambda(b+d)$ which was derived as a conic combination of $ax \geq b$ and $cx \geq d$. At the current node $v$ in the $\SP$ refutation, query $(ax \leq b-1,~ ax \geq b)$. On the branch labelled with $ax \geq b$, query $(cx \leq d-1,~cx \geq d)$. This sequence of queries results in three leaf nodes; see \autoref{fig:SPtCP}. Let the leaf of the branch labelled with $ax \leq b-1$ be $\ell_1$ and let $T_1$ be the subtree rooted at the child of the root of $T$ labelled with $ax \geq b$; recurse on $(T_1,\ell_1)$. Similarly, for the leaf $\ell_2$ of the branch labelled with $cx \leq d-1$, let $T_2$ be the sub-tree rooted at the child of the root of $T$ labelled with $cx \geq d$, and recurse on $(T_2,\ell_2)$. 
		
		For the final leaf, obtained by traversing the edges labelled with $ax \geq b$ and $cx \geq d$, we can derive $0 \geq 1$. To see this, first observe that if $T=G$ (i.e. the base case) then the root node of $T$ is labelled with $0 \geq 1$ and $ax \geq b$ and $cx \geq d$ are the premises used to derive it by a conic combination. In this case, we can derive $0 \geq 1$ by the same conic combination in $\SP$.  Otherwise, by the invariant, the edge leading to $v$ is labelled with the inequality $\lambda(a+c)x \leq \lambda(b+d)-1$. Therefore, a conic combination of this inequality with $ax \geq b$ and $cx \geq d$ yields $0 \geq 1$. 
		
		\item \emph{Division.} If the root of $T$ is labelled with an inequality $ax \geq \lceil b/\delta\rceil$ obtained by division from $\delta ax \geq b$, then query $(\delta ax \leq b-1,~ \delta ax \geq b)$. At the leaf $\ell_1$ corresponding to the edge $\delta ax \leq b-1$, let $T_1$ be the subtree of $T$ rooted at the child of the root of $T$ and recurse on $(T_1,\ell_1)$. At the leaf corresponding to the edge $\delta ax \geq b$ we derive $0 \geq 1$ by a conic combination with  $ax \leq \lceil b/\delta \rceil-1$, which we have already deduced by the invariant. To see this, observe that $b - \delta (\lceil b/\delta \rceil - 1) > 0$
		
		\item \emph{Axiom.}	If $T$ is a single node --- a leaf of the $\tCP$ refutation labelled with some initial inequality $ax \geq b$ of the system that it is refuting --- then, by the invariant, we have already deduced $ax \leq b-1$ and this can be added to $ax \geq b$ to derive $0 \geq 1$. 
	\end{itemize}
 
 	To see that the $\SP$ refutation that we have constructed has depth at most twice that of the $\tCP$ refutation, observe that conic combinations are the only inference rule of $\CP$ for which this construction requires depth $2$ to simulate, while all other rules require depth $1$. 
 	
 	To measure the size, note that every $\CP$ rule with a single premise is simulated in $\SP$ by a single query, where one of the outgoing edges of that query is immediately labelled with $0 \geq 1$.  Each rule with two premises is simulated by two queries in the $\SP$ refutation, where one of the three outgoing edges is labelled with $0 \geq 1$. Therefore, the size of the $\SP$ refutation is $O(s)$.
\end{proof}

%% file: balancingtCPtoSP.tex
A proof system can be \emph{balanced} if any proof of size $s$ implies one of simultaneous size $\poly(s)$ depth $\polylog (s)$.
While it is known that $\tCP$ refutations cannot be balanced, we show next that if we permit the resulting refutation to be in $\SP$, then we can balance. This establishes (ii).

\begin{thm}
\label{thm:balanceSP}
	Any size $s$ $\tCP$ refutation of an unsatisfiable system of linear inequalities $Ax \geq b$ implies a size $O(s)$ and depth $O(\log s)$ $\SP$ refutation of $Ax \geq b$.
\end{thm}
\begin{proof}
	Consider a $\tCP$ refutation of $Ax \geq b$ and let $T$ be its corresponding tree. As well, let $|T|$ denote the number of nodes in $T$. We will construct the $\SP$ refutation recursively; at each step we will keep track of a current node $u$ in the $\SP$ proof we are constructing. The base case is when $|T|=O(1)$, in which case we can use  \autoref{thm:SPsimCPdepth} to create an $\SP$ refutation of $Ax \geq b$ satisfying these properties, and append it to $u$. 
	
		\begin{figure}	
		\centering

		 \begin{tikzpicture}[scale=1]

		\draw[color=black!60, very thick,fill=green!9] (-3,0) -- (-0.3,0) -- (1.5,2) --(0.5,3.5) --cycle ;
		
		\draw[color=black!60, very thick,fill=red!15] (-0.1,-0.3) --(1.6,1.7) -- (2.9,-0.3) -- cycle ;
		
		\draw[very thick, color=black!60] (1.6,1.7)  -- (1.5,2);
		
		\filldraw[color=black!60, fill=betterYellow!25,thick, very thick] (1.6,1.6) circle (6pt);
		
		\node[text width=2cm] at (2.3,0.4) {$T_v$};
		\node[text width=2cm] at (0.3,1.5) {$T \setminus T_v$};
		\node[text width=2cm] at (2.53,1.6) {\scriptsize$v$};
		
		 \end{tikzpicture}	
		 \begin{tikzpicture}[scale=1]

		\draw[thick, ->,color=black!60] (-0.5,0)  -- (-1.45,-0.95);
		
		
		\draw[thick, dashed, ->,color=black!60] (-0.5,-2.3)  -- (-0.1,-2.74);
		\draw[thick, dashed, ->,color=black!60] (-0.5,-2.3)  -- (-0.9,-2.74);
		
		\draw[thick, dashed, ->,color=black!60] (-2.8,-2.3)  -- (-2.4,-2.74);
		\draw[thick, dashed, ->,color=black!60] (-2.8,-2.3)  -- (-3.2,-2.74);

		\draw[thick, ->,color=black!60] (-1.65,-1.1)  -- (-0.68,-2.05);
		\draw[thick, ->,color=black!60] (-1.65,-1.1)  -- (-2.62,-2.05);
		\node[text width=2cm] at (-2.6,-1.5) {\scriptsize$cx \leq d-1$};
		\node[text width=2cm] at (0,-1.5) {\scriptsize$cx \geq d$};
		\filldraw[color=black!60,fill=white,very thick] (-1.65,-1.15) circle (7pt);

		\filldraw[color=black!60,fill=green!9, very thick] (-0.5,-2.25) circle (7pt);

		\filldraw[color=black!60,fill=red!15,very thick] (-2.78,-2.28) circle (7pt);
		
		\node[text width=2cm] at (0,-3.1) {$T \setminus T_v$};
		\node[text width=2cm] at (-2,-3.1) {$T_v$};
		\node[text width=2cm] at (0.35,-2.26) {\scriptsize$u_2$};
		\node[text width=2cm] at (-1.92,-2.26) {\scriptsize$u_1$};
		\node[text width=2cm] at (-0.74,-1.15) {\scriptsize$u$};

		 \end{tikzpicture}	
		 \caption{A decomposition of $\tRCP$ tree $T$ into $T_v$ and $T \setminus T_v$ (left) and the corresponding partial $\SP$ refutation (right).}
		 \label{fig:balancetCP}
	\end{figure}
	
	For the recursive step,  observe that because the tree has fanin at most $2$, there exists a node $v$ in $T$ such that the subtree $T_v$ rooted at $v$ satisfies $|T|/3 \leq |T_v| \leq 2|T|/3$. Let $cx \geq d$ be the line corresponding to $v$. At node $u$ in the $\SP$ proof, query $(cx \leq d-1,~ cx \geq d)$. Let $u_1$ (resp. $u_2$) be the child of $u$ obtained by following the edge labelled with $cx \leq d-1$ (resp. $cx \geq d$); see \autoref{fig:balancetCP}. We recurse as follows: 
	\begin{itemize}
		\item At $u_1$; Observe that the sub-proof $T_v$ is a $\tCP$ derivation of the inequality $cx \geq d$. Because we have deduced $cx \leq d-1$ on the path to $u_1$, if we also deduce $cx \geq d$ then this is sufficient to derive $0 \geq 1$. Therefore, at $u_1$ we recurse on the $\tCP$ derivation $T_v$.
		\item At $u_2$: Observe that the sub-proof $T \setminus T_v$ is a $\tCP$ refutation of $Ax \geq b$ where we have assumed $cx \geq d$ as an axiom. Therefore, at $u_2$ we recursively construct an $\SP$ refutation of the set of inequalities $\{Ax \geq b, cx \geq d\}$ using tree $T\setminus T_v$.
	\end{itemize} 
	The size of the $\tCP$ refutation is clearly preserved. Observe that the depth of the resulting $\SP$ refutation becomes logarithmic in $s$, since we are reducing the size of the proof to be simulated by a constant factor on each branch of a query.	
\end{proof}

%% file: SpaceCPtoSP.tex
Next,we show how to balance $\CP$ proofs into $\SP$, provided the \emph{space} at each step of the proof is bounded. 

The space for a proof system models the amount of information that must be remembered at each state in the nondeterministic Turing machine that underlies a proof system. To capture this, we redefine a Cutting Planes refutation of a system of linear inequalities $Ax \geq b$ as a sequence of configurations $C_1,\ldots, C_s$ where a configuration $C_i$ is a set of integer linear inequalities satisfying the following conditions: (i) $C_1 = \emptyset$, (ii) $C_s$ contains the inequality $0 \geq 1$, (iii) each configuration $C_t$ follows from $C_{t-1}$ by removing any number of inequalities and including an inequality which was derived from inequalities in $C_{t-1}$ by one of the rules of $\CP$ or an initial inequality belonging to $Ax \geq b$. The \emph{line space} of a refutation is $\max_{i \in [s]} |C_i|$, the maximum number of inequalities in any configuration. 

\begin{thm}
	For any $\CP$ refutation of size $s$ and line space $\ell$ of a system of linear inequalities $Ax \geq b$ there is an $\SP$ refutation of depth $O(\ell \log s)$ and size $O(s \cdot 2^\ell)$.
\end{thm}

This implies that strong depth lower bounds on $\SP$ proofs can lead to size-depth tradeoffs for $\CP$.

\begin{proof}
	Fix a Cutting Planes refutation $C_1,\ldots, C_\ell$ where $|C_i| \leq \ell$ for all $i \in [s]$.	The theorem will follow by taking $i=s$ in the following claim.

	\spcnoindent
	{\bf Claim.} For any $i \in [s]$ there exists an $\SP$ tree of depth $2 \ell \log i$ such that every root-to-leaf path ends in a leaf labelled with $0 \geq 1$, except for one, along which we have deduced all of the inequalities in $C_i$. 

\spcnoindent
\emph{Proof of Claim.}	
	It remains to prove the claim. If $C_i$ contains only a single inequality $a_ix \geq b_i$ and it belongs to $Ax \geq b$, then take the tree to be the one corresponding to the $\SP$ query $(a_ix \leq b_i-1,~ a_ix \geq b_i)$. Otherwise, the $\SP$ tree begins with a complete binary tree in which every inequality in $C_{\lfloor i/2 \rfloor}$ is queried. Exactly one path in this tree is labelled with the inequalities in $C_{\lfloor i/2 \rfloor}$, and the remaining paths contain the integer negation (i.e., $cx \leq d-1$) of at least one inequality $cx \geq d$ in $C_{\lfloor i/2 \rfloor}$. We consider these two cases separately.
	
	In the case that a path contains a negation of a line from $C_{\lfloor i/2 \rfloor}$, we attach to its leaf the
$\SP$ tree we obtain recursively by running our construction on $C_1,\ldots, C_{\lfloor i/2 \rfloor}$. The leaves of the resulting tree are all labelled with $0 \geq 1$, except for one. By construction, at the leaf not labelled with $0 \geq 1$ we have deduced all inequalities in $C_{\lfloor i/2 \rfloor}$. Since we attached this tree to a path along which we had deduced the negation of a line in $C_{\lfloor i/2 \rfloor}$, we can label this leaf with a conic combination of these inequalities equalling $0 \geq 1$. The overall depth in this case is $\ell$ for the initial tree and $2\ell\log(\lfloor i/2 \rfloor)$ for the tree obtained recursively. Altogether, $\ell + 2\ell\log(\lfloor i/2 \rfloor) \leq \ell( 1 + 2\log (i) - 2) \leq 2\ell \log i$. 

For the path labelled with the inequalities in $C_{\lfloor i/2 \rfloor}$, note that $C_{\lfloor i/2 \rfloor+1}, \ldots, C_i$ can viewed as configurations of a refutation of the original inequalities $Ax \geq b$ together with the inequalities in $C_{\lfloor i/2 \rfloor}$. At the leaf of this path we have deduced all inequalities in $C_{\lfloor i/2 \rfloor}$. Thus, we can apply the recursive construction to $C_{\lfloor i/2 \rfloor+1}, \ldots, C_i$ to refute this leaf. The overall depth is $\ell + 2\ell\log(\lfloor i/2 \rfloor) \leq \ell (1+2\log(i+2) -2) \leq 2\ell \log i$.
	\begin{figure}	
		\centering
		 \begin{tikzpicture}[scale=1]

		\draw[thick, ->,color=black!60] (-1.65,-1.1)  -- (-0.2,-2.1);
		\draw[thick, ->,color=black!60] (-1.65,-1.1)  -- (-3,-2.1);
		\filldraw[color=black!60,fill=betterYellow!25,thick] (-1.65,-1.15) circle (6pt);

		\draw[color=black!60, thick,fill=green!9] (-1.2,-4) --(0,-2.35) -- (1.2,-4) -- cycle;
		\draw[color=black!60, thick,fill=red!15] (-4.35,-4) --(-3.15,-2.35) -- (-1.95,-4) -- cycle;
		\filldraw[color=black!60, fill=green!9,thick,thick] (-.05,-2.25) circle (6pt);

		\filldraw[color=black!60, fill=red!15,thick] (-3.13,-2.28) circle (6pt);
		
		\node[text width=2cm] at (-5.1,-3.2) {\scriptsize$C_{\lfloor i/2 \rfloor+1}, \ldots, C_i$};
		\node[text width=2cm] at (1.8,-3.2) {\scriptsize$C_1,\ldots, C_{\lfloor i/2 \rfloor}$};
		
		\node[text width=2cm] at (-3,-1.5) {\scriptsize$cx \leq d-1$};
		\node[text width=2cm] at (0.4,-1.5) {\scriptsize$cx \geq d$};
		 \end{tikzpicture}	
		 \caption{The $\SP$ tree corresponding to a configuration $C_i = \{cx \geq d \}$.}
		 \label{fig:spaceDepth}
	\end{figure}
\end{proof}

%% file: non_cg_cuts.tex
So far, we have focused on the relationship between Stabbing Planes and Chv\'{a}tal-Gomory cutting planes. In this section we discuss the relationship between $\SP$ and other popular types of cutting planes. First, we cover the result of Basu et al.~\cite{BasuCSJ20} which shows that $\SP$ can simulate \emph{split cuts}.
Split cuts, which were introduced by Cook et al.~\cite{CookKS90}, and form one of the most popular classes of cutting planes in practical integer linear programming. Recall that an inequality $ax \geq b$ is valid for a polytope $P$ if for every $x^* \in P$, $ax^* \geq b$.
\begin{splitCut}
	A \emph{split cut} for a polytope $P$ is any integer-linear inequality $ax \geq b$ for which there exists a \emph{witnessing} pair $c \in \mathbb{Z}^n$ and $d \in \mathbb{Z}$ such that $ax \geq b$ is valid for both $P \cap \{x \in \mathbb{R}^n : cx \leq d-1\}$ and $P \cap \{x \in \mathbb{R}^n : cx \geq d\}$.
\end{splitCut}
\begin{figure}[h]
\begin{center}
	\begin{tikzpicture}[scale=0.9]
	\filldraw[color=black!60, fill=green!9, very thick] (6.4,-0.5)-- (3.8,0.7) -- (3.9,3.2) -- (6.3,4.5) -- (8.8,3.2) -- (8.7,0.5) -- cycle;
	\filldraw[color=white, fill=white!, very thick,opacity=0.7] (9,4)  -- (3.5,3.9)  -- (3.5,5) -- (9,5) -- cycle;
	\draw[color=black!60, very thick] (5.5,5.3)  -- (4.5,-1.5);
	\draw[color=black!60, very thick,->] (5,2)  -- (4.7,2.04);
	\draw[color=black!60, very thick] (7.5,5.4)  -- (6.5,-1.4);
	\draw[color=black!60, very thick,->] (6.95,1.8)  -- (7.28,1.76);
	
	\draw[color=red!50, very thick,->] (6.3,3.95)  -- (6.3,3.6);
	
	\draw[color=red!50, very thick] (9.,4)  -- (3.5,3.9);
	\node[text width=3cm] at (4.2,2.05) {\small $cx \leq d-1$};
	\node[text width=3cm] at (9,1.75) {\small $cx \geq d$};
	
	\node[text width=3cm] at (7.2,3.3) {\small $ax \geq b$};

	\end{tikzpicture}
	\end{center}
\caption{A split cut $ax \geq b$ witnessed by $(cx \leq d-1,~cx \geq d)$ on the polytope in green.}
\label{fig:split_cut}	
\end{figure}

Split cuts are known to be equivalent to \emph{mixed-integer rounding}  (MIR) cuts \cite{NemhauserW90} and \emph{Gomory mixed integer} cuts \cite{Gomory60}, and generalize \emph{lift-and-project} cuts \cite{BalasCC93}. As well, Dash~\cite{Dash08} gave an example on which split cuts are exponentially separated from Chv\'{a}tal-Gomory cuts and lift-and-project cuts. 
Basu et al.~\cite{BasuCSJ20} showed that split cuts can be simulated in Stabbing Planes. For completeness, we include a proof of this.
\begin{lem}
	[\cite{BasuCSJ20}]	Let $P$ be a polytope and let $P' = P \cap \{ax \geq b\}$ be obtained by a split cut from $P$. Then, there is an $\SP$ tree of size $O(1)$ beginning from $P$ such that every leaf is empty, except for one whose corresponding polytope is $P'$. 
	\end{lem}
	\label{lem:split_cut}
	 \begin{proof}
 	 We simulate the deduction of $P'$ from 
 $P$ in Stabbing Planes as follows:
 		\begin{itemize}
 			\item[(i)] Query $(ax \leq b-1,~ax \geq b)$
 			\item[(ii)] On the branch labelled with $ax \leq b-1$, query $(cx \leq d-1,~cx \geq b)$. Observe that because $ax \geq b$ was valid for both $P \cap \{cx \geq d\}$ and $P \cap \{cx \leq d-1\}$, it follows that both $P \cap \{cx \geq d\} \cap \{ax \leq b-1\}$ and $P \cap \{cx \geq d \} \cap \{ax \leq b-1\}$ are empty. 
 		\end{itemize}
 		Therefore, the only non-empty leaf is the one corresponding to $P'$.
 		\end{proof}
 Dash~\cite{Dash08} studied split cuts as a proof system, and showed that the lower bound of Pudlak~\cite{Pudlak97} could be extended to prove exponential lower bounds on the length of split cut proofs. A split cut refutation of a system of integer linear inequalities $Ax \geq b$ (representing a polytope $P$) is a sequence of inequalities $c_1x \geq d_1, \ldots, c_s x \geq d_s = 0 \geq 1$ such that $c_ix \geq d_i$ is a split cut for the polytope $P \cap \{c_jx \geq d_j\}_{j <i}$. 
 The following is immediate corollary of \autoref{lem:split_cut}.
 
 \begin{cor}
 	Stabbing Planes polynomially simulates split cut proofs.	
 \end{cor} 
 \begin{proof}
 		To simulate any split cut refutation $c_1x \geq b_1,\ldots, c_sx \geq b_s$, we simply simulate each cut inductively using \autoref{lem:split_cut}.
 \end{proof}

Finally, we note that there exist cutting planes that cannot be efficiently simulated by $\SP$. This is witnessed by the fact that $\SP$ cannot polynomially simulate \emph{semantic} $\CP$ \cite{FilmusHL16}. A concrete example of a type of cutting plane that $\SP$ likely cannot simulate are the \emph{matrix cuts} of Lov\'{a}sz and Schrijver \cite{LovaszS91}. As we describe in \autoref{sec:SP_relations_to_other_proofs} $\CP$, and therefore, $\SP^*$ cannot quasi-polynomially simulate the Lovasz-Schrijver proof system. However, whether this holds for $\SP$ with unbounded coefficients remains an interesting question.

%% file: SPEqualstRCP.tex
The Resolution over Cutting Planes ($\RCP$) proof system was introduced by Kraj\'{i}\v{c}ek \cite{Krajicek98} as a mutual generalization of both Cutting Planes and resolution --- the lines of an $\RCP$ proof are clauses of integer-linear inequalities, and in a single
step one can take two previously derived disjunctions and either apply a Cutting Planes rule to a single inequality in the disjunctions, or apply a resolution-style ``cut''.

\begin{Reskdef} An $\RCP$ \emph{proof} of a disjunction $\Gamma_s$  from a system of integer-linear inequalities $Ax \geq b$ is a sequence of disjunctions $P = \{\Gamma_i\}_{i \in [s]}$ such that each $\Gamma_i$ is a disjunction which is either an inequality from $Ax \geq b$ or was derived from earlier disjunctions by one of the following deduction rules: 

\begin{itemize}
	\item \emph{Conic Combination}. From $(ax \geq b) \vee \Gamma$ and $(cx \geq d) \vee \Gamma$ deduce $(\lambda (a+c)x \geq \lambda (b+d)) \vee \Gamma$ for any non-negative integer $\lambda$. 
	\item \emph{Division}. From $(ax \geq b) \vee \Gamma$ and integer $\delta$ dividing each entry of $a$, deduce $((a/\delta)x \geq \lceil b/\delta \rceil) \vee \Gamma$.
	\item \emph{Cut}. From $(ax \geq b) \vee \Gamma$ and $(ax \leq b-1) \vee \Gamma$ derive $\Gamma$.
	\item \emph{Weakening}. From $\Gamma$ deduce $\Gamma \vee (ax \geq b)$
	\item \emph{Axiom Introduction}.  Deduce $(ax \geq b) \vee (ax \leq b-1)$ for any integer-linear inequality $ax \geq b$.
	\item \emph{Elimination}. From $(0 \geq 1) \vee \Gamma$ deduce $\Gamma$.
\end{itemize}
 The \emph{size} of a proof is the number of disjunctions $s$ in the proof 
  and the \emph{width} of the proof is the maximal number of inequalities in any disjunction in the proof.
 An $\RCP$ \emph{refutation} of $Ax \geq b$ is a proof of the empty clause $\Lambda$ from $Ax \geq b$.
  The proof system $\tRCP$ is the tree-like restriction of $\RCP$ in which the underlying implication graph is required to be a tree.
  \end{Reskdef}
  
  The main result of this sub-section is that $\SP$ is polynomially equivalent to $\tRCP$. 
  \SPeqtRCP*

  Even though $\SP$ turns out to be equivalent to a system already in the literature, this new perspective has already shown to be useful: none of aforementioned results were known for $\tRCP$. 
  
  We will prove \autoref{thm:SPEqualstRCP} in two parts. 
\begin{claim}
	Let $Ax \geq b$ be an unsatisfiable system of $m$ integer-linear inequalities. Any size $s$ and depth $d$ $\SP$ refutation implies a $\tRCP$ refutation of size $O(s(d^2+dm))$ and width $d+1$.
\end{claim}
\begin{proof}
	Consider an $\SP$ refutation of $Ax \geq b$ of size $s$ and depth $d$. Fix any root-to-leaf path $p$ in the refutation and let $c_1x \geq d_1, \ldots, c_tx \geq d_t$ be the sequence of linear inequalities labelling $p$. We will first show how to derive the clause
	\begin{align} (c_1x \leq d_1-1) \vee \ldots \vee (c_tx \leq d_t-1) \label{eq:firstLineSP=tRCP} \end{align} 
	in $\tRCP$. For every $i \in [t]$, using \emph{axiom introduction}, introduce $(c_i x \leq d_i -1) \vee (c_ix \geq d_i)$ and \emph{weaken} it to obtain 
	\begin{align}(c_ix \geq d_i) \vee (c_1x \leq d_1-1) \vee \ldots \vee (c_tx \leq d_t-1) . \label{eq:secondLineSP=tRCP}
	\end{align} 
	As well, weaken each initial inequality $a_ix \geq b_i$ in $Ax \geq b$ to 
	\begin{align} (a_i x \geq b_i )\vee (c_1x \leq d_1-1) \vee \ldots \vee (c_tx \leq d_t-1) \label{eq:thirdLineSP=tRCP}
	\end{align} 
	Because $p$ is a root-to-leaf path in the $\SP$ proof, it is labelled with a conic combination of $Ax \geq b$ and $c_i x \geq d_i$ for every $i \in [t]$ equalling $0 \geq 1$. 
	By taking this conic combination of the first inequalities of the lines in (\ref{eq:secondLineSP=tRCP}) and (\ref{eq:thirdLineSP=tRCP}) we can deduce 
	\[ (0 \geq 1 ) \vee (c_1x \leq d_1-1) \vee \ldots \vee (c_tx \leq d_t-1),\] 
	from which we can obtain (\ref{eq:firstLineSP=tRCP}) by  \emph{elimination}. 
	
	Repeat this process to deduce (\ref{eq:firstLineSP=tRCP}) for every root-to-leaf path in the $\SP$ proof. Applying the \emph{cut} rule appropriately to these inequalities yields the empty clause. To see this, let $p$ and $p'$ be two root-to-leaf paths which differ only on their leaf nodes. Then, their corresponding inequalities (\ref{eq:firstLineSP=tRCP}) are of the form
	\begin{align*} &(c_ix \geq d_i) \vee (c_1x \leq d_1-1) \vee  \ldots \vee (c_{t-1}x \leq d_{t-1} -1) \vee (c_tx \leq d_t-1), \\
	&(c_ix \geq d_i) \vee (c_1x \leq d_1-1) \vee \ldots \vee (c_{t-1}x \leq d_{t-1} -1) \vee (c_tx \geq d_t).
	 \end{align*}
	 That is, they differ in their final inequality. Applying the \emph{cut} rule, we can deduce 
	 \[ (c_ix \geq d_i) \vee (c_1x \leq d_1-1) \vee \ldots \vee (c_{t-1}x \leq d_{t-1} -1). \]
	Therefore, by repeating this process we can derive the empty clause.  

	Each deduction of a clause (\ref{eq:firstLineSP=tRCP}) can be done in size $O(t^2+tm+t+m) = O(d^2+dm)$ and has width at most $d+1$. Thus, the size of the proof is at most $O(s(d^2+dm))$. 
  \end{proof}
  We now prove the converse. 
  \begin{claim}
  		Let $Ax \geq b$ be an unsatisfiable system of $m$ integer-linear inequalities. If there is a $\tRCP$ proof of the line $(a_1x \leq b_1-1) \vee \ldots \vee (a_mx \leq b_m -1)$, where $a_i x \geq b_i$ is the $i$th row of $Ax \geq b$, of size $s$ and depth $d$ then there is an $\SP$ refutation of $Ax \geq b$ of size $O(s)$ and depth $2d$. 
  \end{claim}
  \begin{proof}
  Fix such a $\tRCP$ proof of the disjunction. For any clause $\Gamma = (c_1x \geq d_1) \vee \ldots (c_mx \geq d_m)$, we will denote by $\neg \Gamma$ the set of inequalities $\{c_1x \leq d_1-1,\ldots, c_mx \leq d_m-1\}$. 
  We will construct the $\SP$ refutation by structural induction, beginning at the leaves of the refutation and proceeding towards the root. At each line $\Gamma$ in the proof, deduced from children $\Gamma_1$ and $\Gamma_2$, we will assume that we have constructed $\SP$ refutations $\neg \Gamma_1$ and $\neg \Gamma_2$ and use them to construct an $\SP$ refutation of $\neg \Gamma$. 
  
  First, consider a leaf of the proof which, by definition, is an \emph{axiom introduction} of $(cx \leq d-1) \vee (cx \geq d)$ for some arbitrary integer-linear inequality $cx \geq d$. We can construct an $\SP$ refutation of $(cx \geq d) \wedge (cx \leq d-1)$ by querying $(cx \leq d-1,~ cx \geq d)$, and then labelling each leaves with the appropriate conic combination equalling $0 \geq 1$. 
  
  Now, let $\Gamma$ be some line in the proof which was derived from earlier lines $\{\Gamma_i\}$, and suppose that we have constructed $\SP$ refutations of $\{\neg \Gamma_i\}$. To construct a refutation of $\neg \Gamma$, we break into cases based on the rule used to derive $\Gamma$.  
  	\begin{itemize}
  		\item \emph{Conic combination.} Let $\Gamma := (\lambda(c_1+c_2)x \geq \lambda (d_1+d_2)) \vee \Delta$, let $\Gamma_1 := (c_1x \geq d_1) \vee \Delta$, and let $\Gamma_2 := (c_2x \geq d_2) \vee \Delta$. We construct an $\SP$ refutation of $\neg \Gamma$ by first querying $(c_1x \leq d_1-1,~ c_1x \geq d_1)$. On the branch labelled with $c_1x \geq d_1$, apply the $\SP$ refutation of $\neg \Gamma_1$. On the branch labelled with $c_1x \geq d_1$, query $(c_2x \leq d_2-1,~ c_2x \geq d_2)$, and use the refutation of $\neg \Gamma_2$ to refute the branch labelled with $c_2 x \geq d_2$. On the remaining branch, where we have deduced $c_1x \leq d_1-1$ and $c_2x \leq d_2-1$, we have that $0 \geq 1$ is a conic combination with $\lambda (c_1+c_2)x \geq \lambda (d_1+d_2)$.
  		\item \emph{Division.} Let $\Gamma := ((c/\delta)x \geq \lceil d/\delta \rceil) \vee \Delta$ and let $\Gamma_1 = (\delta cx \geq d) \vee \Delta$. Query $(\delta cx \leq d-1,~ \delta cx \geq d)$. On the branch labelled with $cx \leq d-1$ we can use the refutation of $\neg \Gamma_1$. On the branch labelled with $cx \geq d$, it is enough to observe that the intersection of $cx \geq d$ and $((c/\delta)x \leq \lceil d/\delta \rceil - 1)$, provided by $\neg \Gamma$, is empty.  
  		\item \emph{Cut.} Suppose $\Gamma := \Delta$ was derived by cutting on $\Gamma_1:= (cx \geq d) \vee \Delta$  and $\Gamma_2 := (cx \leq d-1) \vee \Delta$. Query $(cx \leq d-1,~cx \geq d)$. On the branch labelled with $cx \leq d-1$ apply the refutation of $\neg \Gamma_1$, and on the branch labelled with $cx \geq d$ use the refutation of $\neg \Gamma_2$.
  		\item \emph{Weakening.} If $\Gamma:= (cx \geq d) \vee \Delta)$ was derived by weakening $\Gamma_1 := \Delta$, then query $(cx \leq d-1,~ cx \geq d)$. On the branch labelled with $cx \leq d-1$ use the refutation of $\neg \Gamma_1$, and the branch labelled with $cx \geq d$ we can deduce $0 \geq 1$ by adding this inequality to $cx \leq d-1$ with is an inequality of $\neg \Gamma$. 
  	\end{itemize}
	Simulating each rule requires at most two queries, of which at most two of the children are not immediately the empty polytope. Therefore, the size of the resulting tree is at most $2s$ and the depth is at most $2d$. 
  \end{proof}

%% file: SPsimResk.tex
Next, we show how to simulate the $k$-DNF resolution proof systems by variants of Stabbing Planes.

\begin{kDNFres}
\label{def:kDNFres}
A $\Res(k)$ refutation of a CNF formula $F$ is a sequence of $k$-DNF formulas $P=\{\Gamma_i\}_{i \in [s]}$ such that $\Gamma_s$ is the empty clause $\Lambda$, and each $\Gamma_i$ is either a clause of $F$ or was derived from earlier DNFs by one of the following deduction rules, where a literal $\ell_i$ is either $x_i$ or $\neg x_i$: 
\begin{itemize} 
	\item \emph{Cut}. From $k$-DNFs $A \lor (\wedge_{i\in S} \ell_i)$ and $B \vee (\lor_{i \in S} \neg \ell)$ deduce $A \lor B$.
	\item \emph{Weakening}. From a $k$-DNF $A$ deduce $A \vee \ell$ for any literal $\ell$. 
	\item \emph{$\wedge$-Introduction}. From $\{A \lor \ell_i \}_{i \in S}$ deduce $A \lor (\land_{i\in S} \ell_i)$.
	\item \emph{$\wedge$-Elimination}. From $A \lor (\land_{i \in S} \ell_i)$ deduce $A \lor \ell_i$ for any $i \in S$.
\end{itemize}
A refutation is \emph{tree-like} if every deduced inequality is used at most once in the refutation (i.e., the underlying implication graph is a tree). The proof system which produces only treelike $\Res(k)$ refutations is denoted $\tRes(k)$.
\end{kDNFres}

\begin{thm}\label{thm:reskSim}
	For any integer $k \geq 1$, any $\Res(k)$ refutation of size $s$ implies an $\RCP$ refutation of size $O(ks)$. Similarly, any $\tRes(k)$ refutation of size $s$ implies an $\SP$ refutation of size $O(ks)$.
\end{thm}

The proof will follow by a straightforward application of the following claim.
\begin{claim}
\label{clm:clausesToInequalities}
	From any disjunction $\bigvee_{i\in S} (x_i \geq 1)\vee \bigvee_{j \in T} (-x_i \geq 0)$, together with inequalities $x_i \geq 0$ and $x_i \leq 1$ for every $i \in [n]$, there is a size $O(|S|+|T|)$ $\tRCP$ derivation of $\sum_{i \in S} x_i + \sum_{j \in T} (1-x_j) \geq 1$. 
\end{claim}
\begin{proof}
	For every $v \in T \cup S$, derive $\sum_{i \in T \setminus \{v\}} x_i + \sum_{j \in S \setminus \{v\}} (1-x_j) \geq 0$ by adding together the inequalities $x_i \geq 0$ and $x_i \leq 1$. For $v \in T \cup S$ add the corresponding inequality to the disjunction in  $\bigvee_{i\in S} (x_i \geq 1)\vee \bigvee_{j \in T} (-x_j \geq 0)$ containing the variable $v$. The result is the disjunction \[\bigvee_{S \cup T} \Big(\sum_{i \in S} x_i + \sum_{j \in T} (1-x_j) \geq 1 \Big),\] which is the inequality $\sum_{i \in S} x_i + \sum_{j \in T} (1-x_j) \geq 1$.  
\end{proof}

\begin{proof}[Proof of \autoref{thm:reskSim}]
	We will show that $\RCP$ can simulate $\Res(k)$. That $\SP$ simulates $\tRes(k)$ will follow by observing that the same proof also shows a simulation of $\tRes(k)$ by $\tRCP$, and then applying \autoref{thm:SPEqualstRCP}.
	
	Let $\{\Gamma_i\}_{i \in [s]}$ be a $\Res(k)$ refutation of a CNF formula $F$, and note that the encoding of $F$ as a system of inequalities (recalled in \autoref{sec:SPdef}) includes $x_i \geq 0$ and $x_i \leq 1$ for every $i \in [n]$. We will encode each disjunction $\Gamma := \Delta_1 \lor \ldots \lor \Delta_t$ as follows: each $\Delta:= (\land_{i\in S} x_i) \land (\land_{j \in T} \neg x_j)$ is represented by the inequality $\sum_{i\in S} (x_i-1) + \sum_{j \in T} -x_j \geq 0$; observe that both representations are satisfied by the same set of $\{0,1\}$-assignments. Let $L_\Gamma$ be the encoding of $\Gamma$ obtained by replacing each $\Delta_i$ by its encoding as an inequality.

It remains to show that $\RCP$ can simulate the deduction rules of $\Res(k)$.
	\begin{itemize}
		\item \emph{Cut}. Suppose that $\Gamma:= A \lor B$ be deduced by cutting on $\Gamma_1 := A \lor (\land_{i \in S} x_i) \wedge (\land_{j \in T} \neg x_j)$ and $\Gamma_2:= B \lor (\lor_{i \in S} \neg x_i) \lor (\lor_{j \in T} x_j)$. As well, suppose that we have already deduced the corresponding lines $L_{\Gamma_1}:= L_A \vee (\sum_{i \in S} (x_i-1) + \sum_{j \in T} -x_j \geq 0)$ and $L_{\Gamma_2} := L_B \vee \bigvee_{i \in S} (-x_i \geq 0)\vee \bigvee_{j \in T} (x_j \geq 1)$.  By \autoref{clm:clausesToInequalities}, $\RCP$ can reencode $L_{\Gamma_2}$ as $L_B \vee (\sum_{i \in S} (1-x_i) + \sum_{j \in T} x_j \geq 1)$, which when added to $L_{\Gamma_1}$ gives $L_A \vee L_B \vee (0 \geq 1)$, which is $L_\Gamma$.
		\item \emph{Weakening}. This is already a rule of $\RCP$.
		\item \emph{$\wedge$-Introduction}. If $\Gamma:= A \lor (\land_{i \in S} x_i) \wedge (\land_{j \in T} \neg x_j)$ was deduced from $\{A \vee x_i\}_{i \in S}$ and $\{A \vee \neg x_j\}_{j \in T}$ and we have already deduced $L_{\Gamma_i}:=L_A \lor (x_i \geq 1)$ and $L_{\Gamma_j}:=L_A \lor (-x_j \geq 0)$ for all $i \in S$ and $j \in T$. Then $L_{\Gamma}$ can be deduced by adding together all of the $L_{\Gamma_i}$ and $L_{\Gamma_j}$.
		\item \emph{$\wedge$-Elimination}. If $\Gamma = A \lor x_i$ was deduced from $\Gamma_1:=A \vee (\wedge_{j \in S} x_j) \land (\wedge_{t \in T} x_t)$, then $L_{\Gamma}$ can be deduced from $L_{\Gamma_1}:= A \lor (\sum_{j \in S} (x_j-1) + \sum_{t \in T} -x_t \geq 0)$ by adding the inequalities $x_j \leq 1$ for every $j \in S\setminus \{i\}$ and $x_t \geq 0$ for every $t \in T$. A similar argument holds if $x_i$ is negated.
	\end{itemize}
\end{proof}

Atserias, Bonet, and Estaban \cite{AtseriasBE02} gave polynomial-size proofs of the \emph{clique-coclique} formulas in $\tRes(k)$, for cliques of size $\Omega(\sqrt{n})$ and cocliques of size $o(\log^2 n)$. For this range of parameters, quasi-polynomial size lower bounds are known \cite{Pudlak97}. This rules out 
the possibility of a \emph{polynomial} simulation of $\RCP$ or $\tRes(k)$ by Cutting Planes.

%% file: SPDepth.tex
In this section we prove \autoref{thm:SPdepthLB}, which we restate next for convenience.

\SPdepthLB*

Note that every unsatisfiable CNF formula has a refutation in depth $n$ by simply querying $(x_i \leq 0,~ x_i \geq 1)$ for all $i \in [n]$. Therefore, this lower bound is tight up to a $\log^2 n$ factor.

The proof proceeds by showing that from any shallow $\SP$ refutation we can extract a short \emph{randomized} or \emph{real} communication protocol for the associated \CNFSearch. The lower bound follows by appealing to known lower bounds on the communication complexity of this problem.

\begin{lem}
\label{lem:SPtoCC}
	Let $Az \geq b$ be an unsatisfiable system of linear equations encoding a CNF formula $F$ and let $(X,Y)$ be any partition of the variables $z$. Every depth $d$ $\SP$ refutation of $Az \geq b$ implies a $O(d\log n + \log ^2 n)$-round  randomized communication protocol and a $O(d + \log n)$-round real communication protocol for solving $\Search_{X,Y}(F)$.
\end{lem}
\begin{proof}
	We will first present a general procedure for solving the false clause search problem and then show how to instantiate it in both models of communication. 
	
	Fix an $\SP$ refutation of $Az \geq b$.
	Let Alice be given a boolean assignment to $X$ and Bob be given a boolean assignment to $Y$. To solve the search problem, they will follow the root-to-leaf path through the refutation, maintaining the invariant that their joint assignment $(X,Y)$ satisfies all of the inequalities labelling the root to leaf path. Suppose that they have arrived at a node in the refutation corresponding to a query $(cz \leq d-1,~ cz \geq d)$. Observe that their joint assignment $(X,Y)$ to $z$ satisfies exactly one of these two inequalities. They will proceed down the path corresponding to the satisfied inequality, thus preserving their invariant. 
	
	Once they arrive at a leaf, they will use the conic combination of inequalities which evaluates to $0 \geq 1$ that labels it in order to search for an inequality of $Az \geq b$ (corresponding to a clause of $F$) which is falsified by $(X,Y)$. Indeed, by the invariant, the only inequalities in this conic combination which could be falsified by $(X,Y)$ are those belonging to $Az \geq b$, and a falsified inequality must exist because $(X,Y)$ falsifies $0 \geq 1$. 
	 Let the conic combination be $\sum_{i \in [\ell]} \alpha_i c_iz \leq \sum_{i \in [\ell]} \alpha_i d$, where $\alpha_i \geq 0$. By Carath\'{e}odory's Theorem (point (ii) in \FarkasLem), we can assume that $\ell \leq n+2$. To find a falsified inequality, we binary search over the conic combination: test whether $\sum_{i=1}^{\ell/2} \alpha_i c_i z \leq \sum_{i=1}^{\ell/2} \alpha_i d_i$ is falsified by $(X,Y)$. If it is, recurse on it; otherwise, recurse on $\sum_{i=\ell/2+1}^{\ell} \alpha_i c_i z \leq \sum_{i=\ell/2+1}^{\ell} \alpha_i d_i$. Because $\ell \leq n+2$, this process terminates in $O(\log n)$ rounds having found an inequality belonging to $Az \geq b$ which is falsified by $(X,Y)$. 
	
	To implement this procedure in communication, it remains to show  that linear inequalities can be evaluated efficiently in each of the models. 
	\begin{itemize}
		\item \emph{Real communication}: this can be done in a single round of communication. If Alice and Bob want to evaluate $c_1x+c_2y \geq d$, then Alice can send $c_1x$ to the referee and Bob can send $d-c_2y$. The referee  returns whether $c_1x \geq d- c_2y$.
		\item \emph{Randomized communication}: this can be done in $O(\log n)$ rounds of communication by combining the following two results. The first is the $O(\log b)$ protocol of Nisan~\cite{Nisan93} for deciding a linear inequality representable in $b$ bits. The second is a result due to Muroga~\cite{Muroga71} which states that for any linear inequality on $n$ variables, there exists a linear inequality whose coefficients are represented in $O(n \log n)$ bits and which has the same output on points in $\{0,1\}^n$.
	\end{itemize}
\end{proof}

To establish \autoref{thm:SPdepthLB}, it remains to lower bound the communication complexity of $\Search(F)$. Strong lower bounds on the randomized communication complexity of the false clause search lower bound were proven by G\"{o}\"{o}s and Pitassi \cite{GoosP18}. In particular, Theorem 8.1 in \cite{GoosP18} gives an unsatisfiable CNF formula $F$ on $\poly(n)$ many clauses and partition $(x,y)$ of the variables for which the randomized commutation complexity of $\Search_{x,y}(F)$ requires $\Omega(n/\log n)$ rounds. Together with \autoref{lem:SPtoCC}, this establishes \autoref{thm:SPdepthLB}.

We remark that the formula provided by G\"{o}\"{o}s and Pitassi is somewhat artificial. It is obtained by \emph{lifting} the Tseitin formulas with a \emph{versatile} gadget. By \autoref{thm:Tseitin}
 we know that the Tseitin formulas have $O(\log^2 n)$-depth $\SP$ refutations, and therefore the hardness of these formulas of G\"{o}\"{o}s and Pitassi is derived from the composition with this gadget.
 It remains an open problem to obtain strong lower bounds on the depth of $\SP$ refutations for more natural families of formulas. Towards this, Dantchev et al.~\cite{DantchevGGM21} were able to establish $\Omega(\log n)$ lower bounds on the depth of $\SP$ refutations of the Tseitin formulas and the Pigeonhole principle via new techniques which take into account the geometric  structure of $\SP$ proofs.

%% file: SPBarriers.tex

Next, we explore whether it is possible to leverage this depth lower bound in order to obtain size bounds. 
Throughout this section, we will heavily make use of results of de Rezende, Nordstr{\"{o}}m, and Vinyals \cite{deRezendeNV21}. They established a \emph{lifting theorem} that translates \emph{decision tree} lower bounds for a function $f: \{0,1\}^n \rightarrow \{0,1\}$ to lower bounds on the real communication complexity of the  \emph{composed function} $f \circ \IND_t^n$, which we define next.
Let $\IND_t: [t] \times \{0,1\}^t \rightarrow \{0,1\}$ be the $t$-bit \emph{index} function mapping $(x,y)$ to $y_x$. The function $f \circ \IND^n_t$ is obtained by replacing each variable of $f$ with a copy of $\IND^n_t$ on new variables. For any function $f$, composing with $\IND_t$ induces a \emph{standard partition}, where Alice is given $x \in [t]^n$ and Bob is given $y \in \{0,1\}^{tn}$.

The \emph{decision tree complexity} of a function $f$ is closely related to the DPLL complexity of refuting an unsatisfiable formula. A decision tree is a binary tree in which: (i) every internal node is labelled by a variable $x_i$ and has two outgoing edges labelled with $0$ and $1$, (ii) the leaves are labelled with either $0$ or $1$. A decision tree computes $f$ if for every $x \in \{0,1\}^n$, the leaf obtained by following the root-to-leaf path which agrees with $x$ is labelled with $f(x)$. The decision tree complexity of $f$, denoted $\DT(f)$, is the minimal depth of any decision tree computing $f$. 

\begin{thm}[de Rezende et al. \cite{deRezendeNV21}]
\label{thm:deRezende}
	The following statements hold:
	\begin{itemize}
		\item For any function $f: \{0,1\}^n \rightarrow \{0,1\}$, the real communication complexity of $f \circ \IND^n_{n^4}$ is at least $\DT(f)$.
		\item There is CNF formula $F$ with $\poly(n)$ many clauses which has $\poly(n)$ size resolution refutation but for which any real communication protocol for $\Search_{x,y}(F)$ requires $\Omega(\sqrt{n^{1/4} \log n})$ rounds, for some partition of the variables. 
	\end{itemize}
\end{thm}

\subsubsection{$\SP$ Proofs Cannot be Balanced} 
As an immediate corollary of \autoref{lem:SPtoCC} and \autoref{thm:deRezende}, we show that $\SP$ proofs cannot be balanced. That is, an $\SP$ refutation of size $s$ does not imply one of size $\poly(s)$ and depth $\poly(\log s)$. Thus, superpolynomial $\SP$ size lower bounds do not immediately follow from depth lower bounds. 

\begin{cor} 
	There exists a CNF formula $F$ which has $\poly(n)$ size $\SP$ refutations but any $\SP$ refutation requires depth $\Omega(n^{1/8}/ \log n)$.
\end{cor}
\begin{proof}
	This follows immediately by combining \autoref{thm:deRezende} with \autoref{lem:SPtoCC} together with the fact that $\SP$ can simulate resolution proofs. 
\end{proof}

\subsubsection{Real Communication Cannot be Balanced}

Unlike the randomized protocols, the real communication protocols that result from \autoref{lem:SPtoCC} preserve the topology of the $\SP$ proof. That is, the \emph{size} --- the number of nodes in the protocol tree --- of the resulting real communication protocol is equivalent, up to a $\poly(n)$ factor, to the size of the $\SP$ refutation. Therefore, while $\SP$ proofs cannot be balanced, one might hope that the resulting real communication protocols could be, and thus size lower bounds could still be obtained from depth bounds. 
This is not without precedent; both deterministic and randomized communication complexity \emph{can} be balanced. Furthermore, although it known that $\tCP$ cannot be balanced,  Impagliazzo et al. \cite{ImpagliazzoPU94} show that $\tCP$ refutations of size $s$ can be balanced into $O(\log s)$-round randomized communication protocols for the false clause search problem. 

Surprisingly, we show that real communication protocols \emph{cannot} be balanced. To do so, we establish the first lower bound on the real communication of the \emph{set disjointness} function, perhaps the most well-studied function in communication complexity, which we define next. Let $\OR_n : \{0,1\}^n \rightarrow \{0,1\}$ be the $n$-bit $\lor$-function and $\AND_2 : \{0,1\}^n \rightarrow \{0,1\}$. Then the set disjointness function, $\DISJ_n:= \OR_n \circ \AND_2^n$, is obtained by replacing each of the $n$ input variables of $\OR_n$ by a copy of $\AND_2$ on new variables. As before, this function induces a \emph{standard partition} where Alice is given one of the two input bits of each $\AND_2$ function, and Bob is given the other.

\begin{thm}
\label{thm:realCCBalanced}
	There is a partition of the variables such that 
	$\DISJ_n$ has a real communication protocol of size $O(n)$, but any real communication protocol requires $\Omega((n\log n)^{1/5})$ rounds of communication. 
\end{thm}
This lower bound was subsequently improved to $\Omega(n/\log^2 n)$ by Chattopadhyay, Lovett, and Vinyals \cite{ChattopadhyayLV19}.

The main technique for obtaining lower bounds on the real communication complexity of a function is by a \emph{lifting theorem}, reducing the task of proving lower bounds on certain \emph{composed functions} to the decision tree complexity of the un-composed function. Although $\DISJ_n$ is a composed function, there is currently no lifting theorem for composition with the $AND_2$ function. We circumvent this by exploiting the fact that $\DISJ_n$ is \emph{complete} for the class $\NP^{cc}$ of functions with polylogarithmic \emph{nondeterministic communication} protocols. 

\begin{np_cc}
The \emph{nondeterministic communication complexity} of a function $f:\{0,1\}^n \rightarrow \{0,1\}$ and a partition $(X,Y)$, is the length of the shortest string $z \in \{0,1\}^\ell$ that can convince Alice and Bob to accept an input $(x,y) \in f^{-1}(1)$ (without communicating). That is, it is the smallest $\ell$ such that for every $(x,y) \in f^{-1}(1)$ there is a string $z \in \{0,1\}^\ell$ such that both Alice and Bob accept, and for every $(x,y) \in f^{-1}(0)$ and every string $z \in \{0,1\}^\ell$, either Alice or Bob rejects.
\end{np_cc}

To prove the lower bound on $\DISJ_n$ we find a function in $\NP^{cc}$ to which known lifting theorems for real communication  can be applied. Then, we use $\NP^{cc}$-completeness to transfer this lower bound to $\DISJ_n$. The function that we will use is $\OR_n \circ \IND_t$. First, we show that this function belongs to $\NP^{cc}$. 

\begin{lem}
\label{lem:NPccProtocolForDisj}
	There is a $O(\log t + \log n)$ $\NP^{cc}$ protocol computing $\OR_n \circ \IND_t^n$ for the standard partition associated with $\IND_t$.
\end{lem}
\begin{proof}
	Fix some input $(x,y) \in [t]^n \times \{0,1\}^{nt}$ and observe that the $i$th input bit to $\OR_n$ can be computed in $\log t+1$ rounds of communication by brute-forcing the index gadget: Alice sends $x_i := x_{i,1},\ldots, x_{i,\log t}$ to Bob who can then compute $\IND_t(x_i,y_i)$, where $y_i := y_{i,1},\ldots, y_{i, t}$, and return the answer to Bob in a single bit.

	\begin{figure}	
		\centering
		 \begin{tikzpicture}[scale=1]
		 
		 \draw[color=black!60, thick,fill=white] (-2,-2) -- (2,-2) -- (2,2) -- (-2,2) -- cycle;
		 \draw[color=black!60, thick,fill=green!9] (-2,-2) -- (-1,-2) -- (-1,-1) -- (-2,-1) -- cycle;
		 \draw[color=black!60, thick,fill=red!15] (2,-2) -- (0,-2) -- (0,0) -- (2,0) -- cycle;
		 \draw[color=black!60,fill=betterYellow!25] (-1,-1) -- (0,-1) -- (0,0) -- (-1,0) -- cycle;
		 \draw[color=red!15,fill=red!15] (-1,-1) -- (0,-1) -- (0,0) -- cycle;
		 \draw[color=black!60, thick] (-1,-1) -- (0,-1) -- (0,0) -- (-1,0) -- cycle;
		 \draw[color=black!60, thick,fill=green!9] (-2,2) -- (-1,2) -- (-1,0) -- (-2,0) -- cycle;
		 \draw[color=black!60, thick,fill=betterYellow!25] (-1,1.5) -- (2,1.5) -- (2,0.5) -- (-1,0.5) -- cycle;
		 \draw[color=black!60, thick,fill=green!9] (1.5,.5) -- (0.5,0.5) -- (0.5,0) -- (1.5,0) -- cycle;
		 
		 \node[text width=1cm] at (-0.8,1.8) {\scriptsize$1$};
		 \node[text width=1cm] at (2.2,1.3) {\scriptsize$2$};
		 \node[text width=1cm] at (1.75,0.3) {\scriptsize$3$};
		 \node[text width=1cm] at (0.25,-.2) {\scriptsize$4$};
		 \node[text width=1cm] at (2.2,-.2) {\scriptsize$5$};
		 \node[text width=1cm] at (-0.8,-1.2) {\scriptsize$6$};
		 
		 \node[text width=2cm] at (-1.5,-0.5) {$x$};
		 \node[text width=2cm] at (0.4,2.4) {$y$};

		 \draw[color=red!85, very thick] (-2,-0.5) -- (2,-0.5);
		 \draw[color=betterYellow!85, very thick] (-0.5,2) -- (-0.5,-2);
		 
		 \node[text width=2cm] at (5,1.5) {$0$};
		 \node[text width=2cm] at (5,1) {$0$};
		 \node[text width=2cm] at (5,0.5) {$0$};
		 \node[text width=2cm] at (5,0) {$1$};
		 \node[text width=2cm] at (5,-0.5) {$1$};
		 \node[text width=2cm] at (5,-1) {$0$};
		 \node[text width=2cm] at (4.7,-2) {$I_A(x)$};
		 
		 \node[text width=2cm] at (6.5,1.5) {$0$};
		 \node[text width=2cm] at (6.5,1) {$1$};
		 \node[text width=2cm] at (6.5,0.5) {$0$};
		 \node[text width=2cm] at (6.5,0) {$1$};
		 \node[text width=2cm] at (6.5,-0.5) {$0$};
		 \node[text width=2cm] at (6.5,-1) {$0$};
		 \node[text width=2cm] at (6.2,-2) {$I_B(y)$};
		 
		 \draw[color=black!60, thick] (4,1.9) -- (3.8,1.9) -- (3.8,-1.4) -- (4,-1.4);
		 \draw[color=black!60, thick] (4.2,1.9) -- (4.4,1.9) -- (4.4,-1.4) -- (4.2,-1.4);
		 
		 \draw[color=black!60, thick] (5.5,1.9) -- (5.3,1.9) -- (5.3,-1.4) -- (5.5,-1.4);
		 \draw[color=black!60, thick] (5.7,1.9) -- (5.9,1.9) -- (5.9,-1.4) -- (5.7,-1.4);
		 \end{tikzpicture}	

		 \caption{A covering of the communication matrix with monochromatic rectangles (left) and the corresponding $\DISJ_n$ instance (right).}
		 \label{fig:reductionToDISJ}
	\end{figure}
	
	Consider the following protocol $\NP^{cc}$ for $\OR_n \circ \IND_t^n$: Alice and Bob are given a $\log n$-bit string encoding the index $i \in [n]$ where $\IND_t(x_i,y_i) = 1$; that is, $i$ witnesses that $(x,y)$ is accepting input of $\OR_n \circ \IND_t^n$; see \autoref{fig:reductionToDISJ}. Alice and Bob verify that indeed $\IND_t(x_i,y_i) = 1$ by performing the above brute force protocol in $\log t$ rounds of communication. 
\end{proof}

\begin{lem}
\label{lem:lowerBoundORIND}
	Let $m= n^4$, then any real communication protocol computing $\OR_n \circ \IND_t^n$ over the standard partition requires $\Omega(n \log n)$ rounds.
\end{lem}

\begin{proof}
	Observe that the  decision tree complexity of computing $\OR_n$ is $n$ (since the \emph{sensitivity} of $OR_n$ is $n$). The proof follows by combining this with \autoref{thm:deRezende}.
\end{proof}

Finally, we are ready to prove the main theorem of this section. 

\begin{proof}[Proof of \autoref{thm:realCCBalanced}]
	First, we prove the lower bound. We will reduce $\OR_n \circ \IND_t^n$ for $t=n^4$ to $\DISJ_n$. By \autoref{lem:NPccProtocolForDisj} there is a cover of the $1$-entries of the communication matrix of $\OR_n \circ \IND_t^n$ by at most $2nt$ monochromatic rectangles. Enumerating this rectangle covering gives us an instance of set disjointness: on input $(x,y)$ to $\OR_n \circ \IND_t^n$, Alice and Bob construct indicator vectors $I_A(x)$ and $I_B(y)$ of the rectangles in this rectangle covering in their respective inputs lie. Then $\OR_n \circ \IND_t^n = 1$ iff $\DISJ(I_A(x), I_B(y)) = 1$. 
	
	This instance of $\DISJ_n$ is on $tn/2$ variables, and therefore \autoref{lem:lowerBoundORIND} implies a lower bound of $\Omega(n \log n)$. Letting $\ell = tn$ be the total number of variables, this is a bound of the form $\ell^{1/5} \log \ell$. 
	
	For the upper bound, we give a real communication protocol for $\DISJ_n = \OR_n \circ \AND_2^n$ that has $O(n)$ nodes. Let $x,y$ be the inputs given to Alice and Bob respectively. Sequentially from $i=1,\ldots, n$, they will solve $x_i \wedge y_i$ by Alice sending $x_i$ to the referee and Bob sending $2-y_i$. If they discover that $x_i \wedge y_i = 0$ then they halt and output $0$, otherwise they continue. 
\end{proof}

%% file: ConclusionSP.tex
We end with a several questions left open by this work. First, let us note that several of the questions posed in the original version of this paper were subsequently resolved by \cite{DadushT20} who exhibited an upper bound (\autoref{prop:coefficientBound}) on the size of the coefficients which occur in Stabbing Planes proofs, and showed that the Tseitin formulas could not provide an exponential separation between Cutting Planes and Stabbing Planes. 
\begin{enumerate}
	\item In this work we showed that there are quasipolynomail-size Stabbing Planes proofs of the Tseitin formulas. Can this be improved to polynomial? 
	\item A recent work \cite{FlemingPR21} exhibited \emph{supercritical} size/depth tradeoffs for Cutting Planes --- exhibiting a formula for which any small proof must have depth which goes far beyond worst-case. This built upon an earlier supercritical size/width tradeoff for tree-like Resolution by Razborov~\cite{BeameBI16,Razborov16}. This is in contrast to sufficiently expressive proof systems, such as $\AC^0$-Frege, which \emph{can} be balanced. Depth captures the degree to proofs --- and therefore algorithms which they formalize --- can be parallelized. Furthermore, the depth in integer-programming based proof systems such as Stabbing Planes is closely related to rank measures of polytopes, which are studied in integer programming theory. In this work, we showed that Stabbing Planes cannot be balanced. We ask whether this can be improved to a supercritical size/depth tradeoff.
	\item We have shown transformations of Cutting Planes proofs into Stabbing Planes which preserve either the size or the depth of the original proof. Does there exist a transformation which preserves both parameters \emph{simultaneously}? 
	\item Establish super-polynomial lower bounds on the size of Stabbing Planes proofs. As mentioned in the related work section, \cite{FlemingGIPRTW21} proved superpolynomial lower bounds on $\SP$ proofs with coefficients of magnitude bounded above by $2^{n^\delta}$ for some constant $\delta >0$ by reducing to Cutting Planes lower bounds, however it is unclear whether this reduction can be made to work for arbitrarily large coefficients. Dadush and Tiwari~\cite{DadushT20} exhibited an upper bound of $\exp(\poly(n))$ on the magnitude of the coefficients in any Stabbing Planes proof. Thus, one potential (although seemingly unlikely) way to resolve this question is to improve their upper bound to $2^{n^\delta}$. Another approach for obtaining lower bounds on general Stabbing Planes proofs, suggested by Garg et al.~\cite{GargGKS18}, would be to obtain lifting theorem for intersection of triangles trees. 
	\item Fleming et al.~\cite{FlemingGIPRTW21} showed that any bounded-weight Stabbing Planes ($\SP^*$) proof can be quasi-polynomially translated into Cutting Planes. Can this simulation be improved to handle $\SP$ proofs of arbitrarily large coefficients? Alternatively, can we separate $\SP$ from $\CP$? 
	\item As mentioned in the introduction, we feel that $\SP$ has potential, in combination with state-of-the-art algorithms for SAT, for improved performance on certain hard instances, or possibly to solve harder problems such as maxSAT or counting satisfying assignments. The upper bound on the Tseitin example illustrates the kind of reasoning that $\SP$ is capable of: arbitrarily splitting the solution space into sub-problems based on some measure of progress. This opens up the space of algorithmic ideas for solvers and should allow one to take fuller advantage of the expressibility of integer linear inequalities. For example, since geometric properties of the rational hull formed by the set of constraints can be determined efficiently, an $\SP$-based solver could branch on linear inequalities representing some geometric properties of the rational hull. Therefore, it is an open problem to realize a $\SP$ based solvers or to implement $\SP$-like branching in conjunction with current solvers.
\end{enumerate}